\documentclass[twocolumn,showpacs,preprintnumbers]{revtex4-1}

\usepackage{epsfig}
\usepackage{graphicx}
\usepackage{epstopdf, epsfig}

\usepackage{amsmath}
\usepackage{amssymb}

\usepackage{natbib}

\usepackage{color}

\newcommand\beq{\begin{equation}}
\newcommand\eeq{\end{equation}}

\newcommand{\ben}{\begin{eqnarray}}
\newcommand{\een}{\end{eqnarray}}
\newcommand{\benn}{\begin{eqnarray*}}
\newcommand{\eenn}{\end{eqnarray*}}

\newcommand{\apar}{ A_{\parallel}}

\newcommand{\bpar}{B_z}
\newcommand{\pa}{\partial}

\newcommand{\Ppe}{P_{\parallel e}}
\newcommand{\Ppee}{P_{\perp e}}
\newcommand{\Qpe}{Q_{\parallel e}}
\newcommand{\Qpee}{Q_{\perp e}}

\newcommand{\Rxe}{R_{\| \perp e}}
\newcommand{\Rpe}{R_{\perp \perp e}}
\newcommand{\Ppi}{P_{\parallel i}}
\newcommand{\Ppei}{P_{\perp i}}
\newcommand{\Qpi}{Q_{\parallel i}}
\newcommand{\Qpei}{Q_{\perp i}}

\newcommand{\Rxi}{R_{\| \perp i}}

\newcommand{\Rpi}{R_{\perp \perp i}}

\newcommand{\ppe}{p_{\parallel e}}
\newcommand{\ppee}{p_{\perp e}}

\newcommand{\ppi}{p_{\parallel i}}
\newcommand{\ppei}{p_{\perp i}}

\newcommand{\Tpei}{T_{\perp i}}

\newcommand{\ds}{\Delta_s}
\newcommand{\lapp}{\Delta_\perp}
\newcommand{\edi}{\mathrm{e}^{\tau \ds}}
\newcommand{\ede}{\mathrm{e}^{\delta^2 \ds}}
\newcommand{\eddi}{\mathrm{e}^{2\tau \ds}}
\newcommand{\edde}{\mathrm{e}^{2\delta^2 \ds}}
\newcommand{\dd}{\delta^2}
\newcommand{\gpar}{\nabla_{\parallel}}

\newcommand{\bhatb}{{\widehat{\boldsymbol b}}}
\newcommand{\bhatz}{{\widehat{\boldsymbol z}}}

\newcommand{\bnabla}{\mathbf{\nabla}}
\newcommand{\bcdot}{\mathbf{\cdot}}

\newcommand{\mathsfbi}{\mathbf}

\newcommand{\nno}{\nonumber}
\newcommand{\bilapp}{\Delta_{\perp}^2}
\newcommand{\gperp}{\nabla_{\perp}}
\newcommand{\tpei}{t_{\perp_i}}
\newcommand{\tpi}{t_{\parallel i}}

\allowdisplaybreaks

\begin{document}

\title{Fluid and gyrofluid modeling of low-$\beta_e$ plasmas: \\ 
	phenomenology of kinetic Alfv\'en wave turbulence}

\author{T. Passot$^1$,  P.L. Sulem$^1$ and E. Tassi$^2$}

\affiliation{$^1$ Universit\'e C\^ote d'Azur, CNRS, Observatoire de la C\^ote d'Azur,
Laboratoire J.L. Lagrange, Boulevard de l'Observatoire, CS  34229, 06304 Nice Cedex 4, France \\
$^2$ Aix Marseille Univ, Univ Toulon, CNRS, CPT, Marseille, France }

\begin{abstract}
Reduced fluid models  including electron inertia and ion finite Larmor radius corrections are derived asymptotically, both from fluid
basic equations and from a gyrofluid model. They apply to
collisionless plasmas with small ion-to-electron equilibrium temperature ratio and low $\beta_e$, where $\beta_e$ indicates the ratio between the equilibrium electron pressure and the magnetic pressure exerted by a strong, constant and uniform magnetic guide field. The consistency between the fluid and gyrofluid approaches is ensured when choosing ion closure relations prescribed by the underlying ordering. A two-field reduction of the gyrofluid model valid for arbitrary equilibrium temperature
ratio is also introduced, and is shown to have a noncanonical Hamiltonian structure. This model provides a convenient
framework for studying kinetic Alfv\'en wave turbulence, from MHD to sub-$d_e$ scales (where $d_e$ holds for the electron skin depth). Magnetic energy spectra are phenomenologically determined within energy and generalized helicity cascades
in the perpendicular spectral plane. Arguments based on absolute statistical equilibria are used to predict the direction of the transfers, pointing out that, within the sub-ion range associated with a $k_\perp^{-7/3}$ transverse magnetic spectrum, the generalized helicity could display an inverse cascade if injected at small scales, for example by reconnection processes.
\end{abstract}


\maketitle
\section{Introduction}

Reduced fluid models including electron inertia are classically used to study collisionless magnetic reconnection. These models, which are limited to scales large with respect to the electron Larmor radius $\rho_e$, require a small value of the electron beta parameter $\beta_e$ defined as the ratio between the equilibrium electron pressure and the magnetic pressure exerted by a strong, constant and uniform magnetic guide field.
Similarly, at the level of the ions, a fluid computation  of ion finite
Larmor radius (FLR) corrections restricts the considered scales to be either much larger than the ion Larmor radius $\rho_i$ (for which a perturbative approach is possible) or much smaller than $\rho_i$, a case studied in Ref. \cite{PST17},
where the ion velocity is negligible. Denoting with $\tau$ a constant equilibrium ion-to-electron temperature ratio, when concentrating on scales of the order of the
sonic Larmor radius $\rho_s$, defined as $\rho_s=\rho_i/\sqrt{2 \tau}$, these regimes correspond 
to a value of $\tau$  much smaller or much larger
than unity, respectively. In the case where magnetic fluctuations along the guide field are retained, the case of negligible $\tau$
was addressed in two dimensions in Refs. \cite{Fitzpatrick04,FPerratum}
and extended to three dimensions in Ref. \cite{Tassi10}.  The case where  $\tau$ is small but not totally negligible (or finite, provided the considered scales are assumed larger than $\rho_i$) was addressed in Ref. \cite{Hsu86}, when electron inertia is neglected.
One of the motivations of the present paper is to extend this
four-field model by retaining electron inertia, using a rigorous asymptotic ordering.
Such a small-$\tau$ asymptotics, performed at scales of the order of the sonic Larmor radius $\rho_s$, involves a second order computation of the ion FLR 
corrections in terms of $k_\perp \rho_i$, where $k_\perp$ refers to the  
transverse wavenumber of the fluctuations. As will be shown, the resulting 
reduced fluid model can also be obtained
as an asymptotic limit of the gyrofluid model derived in  Ref. \cite{Bri92}. The question then arises
of the consistency of the two approaches, an issue which may be sensitive to the closure assumptions. The case of finite  $\tau$ can be addressed using a gyrofluid approach which,
retaining the parallel magnetic fluctuations $B_z$, remains valid for somewhat
larger values of $\beta_e$, at least at large enough scales. When  reduced to two fields by  neglecting the coupling to the parallel ion velocity $u_i$ and thus to the slow magneto-acoustic modes, the resulting  gyrofluid model isolates the dynamics of kinetic Alfv\'en waves (KAWs) which are supposed to play a main role in 
the solar wind.

Another aim of this paper is to use this two-field gyrofluid model to 
study phenomenologically critically-balanced KAW turbulence 
at scales ranging from MHD to sub-$d_e$ scales (where $d_e$
stands for the electron skin depth), paying a special attention to the 
transverse magnetic energy spectra in the energy or the  generalized helicity
cascades, and to the direct or inverse character of these cascades.
Such Kolmogorov-like phenomenology dismisses the possible effect of 
coherent structures such as current sheets which form
as the result of  the turbulent MHD cascade and which, 
in some instances, can be 
 destabilized by magnetic reconnection. Recent 
 two-dimensional hybrid-kinetic simulations \cite{CerriNJP17} suggest that,
 in the non-collisional regime, this process is fast enough to compete
 with the wave mode interactions, in a way that could affect the 
 cascade at scales comparable to the ion inertial length $d_i$,
 typical of the current sheet width.
 In a small $\beta_i$ plasma, where $\beta_i = \tau \beta_e$, this scale is significantly larger than 
  $\rho_s$, and the spectral break can indeed take
 place at $d_i$, as suggested by recent two-dimensional hybrid
 simulations \cite{Franci16}.
 The above gyrofluid  model can provide an efficient tool to  address this issue.
 
 At this point, it is useful to order the various relevant 
 scales estimated in a homogeneous equilibrium state
 characterized by a  density $n_0$, isotropic ion and electron temperatures $T_{0i}$ 
 and $T_{0e}$, and subject to a strong ambient magnetic field of amplitude $B_0$
 along the $z$-direction. In terms of the  
 sonic Larmor radius $\rho_s  = c_s/\Omega_i$, where $c_s = \sqrt {T_{0e}/m_i}$ is the sound speed
 and $\Omega_i= eB_0/(mc)$ the ion gyrofrequency, one has
 \begin{eqnarray}
 && d_i = \sqrt{\frac{2}{\beta_e}} \rho_s , \qquad
 d_e= \sqrt{\frac{2}{\beta_e}}\delta \rho_s ,\nonumber \\
 && \rho_i = \sqrt{2\tau} \rho_s , \qquad  \rho_e= \sqrt{2} \delta \rho_s,
 \end{eqnarray}
 where $\beta_e= 8 \pi n_0 T_{0e}/B_0^2$, $\delta^2= m_e/m_i$ is
  the electron to ion mass ratio and $\tau = T_{0i}/T_{0e}$. We have here defined the particle Larmor radii ($r=i$ for the ions, $r=e$ for the electrons) by $\rho_r =v_{th\,r}/\Omega_r$ where the particle thermal velocities are given by 
  $v_{th \, r}=(2T_r/m_r)^{1/2}$ and the inertial lengths by $d_r=v_A/\Omega_r$ where $v_A=B_0/(4\pi n_0 m_i)^{1/2}
 =c_s \sqrt{2/\beta_e}$ is the Alfv\'en velocity.
 
 The models to be derived should cover a spectral
 range which includes both scales large compared to 
 $d_i$ (typical of the width of the generated
 current sheets)  and scales comparable to $d_e$ (typical of 
 collisionless reconnection processes). 
 The considered scales will also be assumed to remain large compared to $\rho_e$, so that electron FLR corrections reduce to the contribution ensuring the gyroviscous cancellation. This in particular implies the condition that  $\rho_e/d_e=\beta_e^{1/2}$ be small enough.
 
On the other side, the ion to electron temperature ratio $\tau$ determines the magnitude of  $\rho_i$ relatively to the considered scales. If $\tau\gg 1$, they are much smaller than $\rho_i$, which makes ion velocities  negligible. This case can be addressed using a fluid model, as shown in Ref. \cite{PST17}. For $\tau\ll 1$, they  are much larger than $\rho_i$, and the problem is also amenable to  a fluid approach with ion FLR corrections estimated perturbatively. This regime is addressed in  Section \ref{sec:fluidsmalltau}. For intermediate values of $\tau$, a gyrofluid approach is required. It is the object of Section \ref{sec:gyro}. In Section 
\ref{sec:KAW-turbulence}, a two-fluid restriction of this model is used for a phenomenological study of critically-balanced kinetic Alfv\'en  wave (KAW) turbulence. Section \ref{sec:conclusion} presents a short summary together with a few comments.

\section{Fluid modeling for small $\tau$}\label{sec:fluidsmalltau}

  Two regimes will be here considered with $\beta_e$ scaling either like $\delta$ (scaling I) or like $\delta^2$ (scaling II). The value of $\tau$ must then be chosen so that $\rho_i$ be small enough compared to $\rho_s$ (taken as the characteristic scale), but also smaller than $d_e$. Since $\rho_i/\rho_s=(2\tau)^{1/2}$ and $\rho_i/d_e=(\tau\beta_e)^{1/2}/\delta$, one should take $\tau=O(\delta^{3/2})$ for scaling I and $\tau=O(\delta)$ for
  scaling II.
 In the case of scaling I $d_i/\rho_s\simeq \rho_s/d_e \simeq d_e/\rho_e \simeq 1/\delta^{1/2}$, and $d_e/\rho_i\simeq 1/\delta^{1/4}$,
 whereas for scaling II, $d_e$ and $\rho_s$ are comparable
 and clearly separated from $d_i$ (by a factor $1/\delta$) and $\rho_i$ (by a factor $1/\delta^{1/2}$).
 
 It is convenient to take the sonic Larmor radius $\rho_s$, the 
 sound speed $c_s$ and the inverse ion gyrofrequency $\Omega_i^{-1}$
 as length, velocity and time units. Using the same nondimensional units as in Ref. \cite{TSP16}, the amplitude of the fluctuations of density $n$ and of the
 electric potential $\varphi$ are controlled by the parameter $\varepsilon \ll 1$,
 as $n\sim \varphi =O(\varepsilon)$.  We assume that at scale $\rho_s$,
 $\partial_t =O(\varepsilon)$ and $\bnabla_\perp \sim O(1)$. We denote by $A_\|$ the parallel
component of the magnetic potential, by  $u_i$ and $u_e$ the parallel ion and electron
velocity respectively and by $B_z$ the longitudinal magnetic field fluctuations.
In the case of scaling I, $A_\|\sim u_i \sim \partial_z =O(\varepsilon \delta^{1/2})$, $ u_e = O(\varepsilon / \delta^{1/2})$,  and $B_z=O(\varepsilon \delta)$ (thus to be retained). Differently, for scaling II, $A_\|\sim u_i \sim \partial_z =O(\varepsilon \delta)$, $ u_e = O(\varepsilon / \delta)$, and $B_z=O(\varepsilon \delta^2)$ (thus negligible).
We furthermore denote by ${\mathsfbi P}_r$ the pressure tensor of the $r$ particle species, given by the sum of a gyrotropic part involving the parallel and perpendicular pressure fluctuations $p_{\| r}$ and $p_{\perp r}$ and of a non-gyrotropic contribution ${\boldsymbol \Pi}_r$. Such scalings lead to the 
derivation of reduced fluid models retaining corrections $O(\tau)$ or $O(\delta)$,
relatively to the leading order.

 The Amp\`ere equation reads
\begin{equation}
\Delta_\perp A_\| = \frac{\beta_e}{2} (u_e-u_i).\label{eq:Ampere}
\end{equation}

Summing the equations satisfied by the ion and electron velocities ${\boldsymbol u}_i$
and ${\boldsymbol u}_e$ leads to
\begin{eqnarray}
&& (1+n)\{\partial_t \left ({\boldsymbol u_i} + \delta^2{\boldsymbol u_e}
\right) + {\boldsymbol u}_i \bcdot \bnabla_\perp {\boldsymbol u_i}  
+ \delta^2{\boldsymbol u}_e \bcdot \bnabla_\perp {\boldsymbol u_e}\} \nonumber \\
&& \qquad + \bnabla_\perp \bcdot \Big (\tau {\mathsfbi P}_i
+ {\mathsfbi P}_e \Big)- \frac{2}{\beta_e} {\boldsymbol J}
{\boldsymbol \times }{\boldsymbol B}=0,
\label{vi}
\end{eqnarray}
which, in the small-amplitude (weakly nonlinear) and quasi-transverse asymptotics, gives
\begin{equation}
\frac{d^i}{dt} u_i+\delta^2 \frac{d^e}{dt} u_e+\nabla_\| (\tau p_{\|i}+p_{\| e}) +\tau \bhatb\bcdot\bnabla\cdot
{\boldsymbol \Pi}_i+\bhatb\bcdot\bnabla\cdot {\boldsymbol \Pi}_e=0 \label{eq:upari}
\end{equation}
for the parallel components. Here, we introduced the parallel derivative
  $\nabla_\| f = \bhatb\bcdot\bnabla f =-[ A_\| , f] +\partial_z f$, with $[f, g] = \bhatz \bcdot (\bnabla f \times \bnabla g) =\partial_x f \partial_y g - \partial_y f \partial_x g$, where $f$ and $g$ refer to scalar functions, and $\bhatz$ to the unit vector along the guide field. Noting that, to leading order,
\begin{equation}
\bhatb \bcdot \bnabla_\perp {\boldsymbol \times}
({\boldsymbol J} {\boldsymbol \times} {\boldsymbol B})
= \nabla_\| J_\|= -\nabla_\| \Delta_\perp A_\|,
\end{equation}
one also gets
\begin{eqnarray}
&&\frac{d^i}{dt} \Delta_\perp \varphi_i+\delta^2 \frac{d^e}{dt} \Delta_\perp \varphi_e + \bhatb\cdot \bnabla_\perp\times\bnabla_\perp \cdot (\tau {\mathsfbi P}_i
+ {\mathsfbi P}_e )\nonumber \\
&&+\frac{2}{\beta_e}\nabla_\| \Delta_\perp A_\| =0 \label{eq:vorticities}
\end{eqnarray}
for the sum of the ion and electron vorticities.
Here $\bhatb$ is the unit vector along the local magnetic field and the convective time derivative $d^r/dt$ stands for 
$\partial_t +[\varphi_r, \cdot]$, where the potentials $\varphi_r$ of the leading order transverse velocities of the r-particle species (${\boldsymbol u}_{\perp r} = \bhatb \times \bnabla \varphi_r$), are given by
$\varphi_i = \varphi + \tau p_{\perp i}- ({\tau}/{2}) \Delta_\perp \varphi_i$ and $\varphi_e = \varphi -p_{\perp e}- ({\delta^2}/{2}) \Delta_\perp \varphi_e$. In these formulas,  the first term is associated with the so-called ${\boldsymbol E}\times {\boldsymbol B}$ drift, the second one to the diamagnetic drift, while the last one originates from the leading order non-gyrotropic pressure contribution. Here and in the rest of the paper, electrons will be taken isothermal, leading to $p_{\|e}=p_{\perp e}=n$.

The equations for the magnetic potential and for the parallel magnetic field component are easily obtained (see  Ref. \cite{PST17}) as 
\begin{eqnarray}
&& \partial_t (A_\| -\delta^2 u_e)+ \nabla_\| (\varphi-n)-\delta^2[\varphi_e,u_e]-\bhatb\cdot\bnabla\cdot{\boldsymbol \Pi}_e=0\label{eq:Apar}\nonumber\\\\
&&\frac{d^e}{dt}(B_z-n-\delta^2\Delta_\perp\varphi_e)-\nabla_\| u_e -\bhatb\cdot \bnabla_\perp\times\bnabla_\perp \cdot (\frac{{\boldsymbol \Pi}_e}{1+n})=0.\label{eq:Bz}\nonumber \\
\end{eqnarray}
The system of governing equations is supplemented by the perpendicular pressure balance, obtained by taking the transverse divergence of the transverse component of Eq. (\ref{vi})
considered to leading order, 
\begin{equation}
\frac{2}{\beta_e}B_z=\frac{\tau}{2}\Delta_\perp \varphi_i-\tau p_{\perp i}-n.\label{eq:pressure-balance}
\end{equation}

\subsection{The case of negligible ion temperature}  \label{sec:notemp}

In the case of scaling I,  the system made up of Eqs. (\ref{eq:Ampere}), (\ref{eq:upari}), (\ref{eq:vorticities})-(\ref{eq:pressure-balance}) greatly simplifies as all the non-gyrotropic pressure components become sub-dominant, except the electronic ones associated with the gyroviscous cancellation. The $\tau$ contributions also drop out, and we obtain, writing $\frac{d}{dt}=\partial_t +[\varphi,\bcdot]$,
\begin{eqnarray}
&&\frac{d}{dt}(1+\frac{2}{\beta_e})B_z-\nabla_\| u_e=0\label{eq:FP1}\\
&&\frac{d}{dt}(A_\| -\delta^2 u_e)+\partial_z\varphi+\frac{2}{\beta_e}\nabla_\| B_z=0\label{eq:FP2}\\
&& \frac{d}{dt}(u_i+\delta^2 u_e)-\frac{2}{\beta_e}\nabla_\| B_z=0\label{eq:FP3}\\
&&\frac{d}{dt} \Delta_\perp\varphi+\frac{2}{\beta_e}\nabla_\|\Delta_\perp A_\|=0 \label{eq:FP4}\\
&& u_e-u_i=\frac{2}{\beta_e}\Delta_\perp A_\|,\label{eq:FP5}
\end{eqnarray}
which is a 3D extension of the model presented in Ref. \cite{comisso12}, when 
taken in the cold-ion limit. Note that in this system,  $B_z=-(\beta_e/2) n$. As mentioned in Ref. \cite{comisso12}, it is easy to verify that, up to terms of order $\delta^2$, and after a simple rescaling, these equations also identify, in the 2D case, to those of Refs. \cite{Fitzpatrick04,FPerratum}. A 3D extension of the latter model was given in Ref. \cite{Tassi10}. Both systems possess a Hamiltonian formulation, with the same Poisson bracket structure. In particular, in the 2D limit, they both possess four infinite families of Casimir invariants, three of which associated with Lagrangian invariants.

When writing the above system using the Alfv\'en velocity instead of the ion sound speed
as velocity unit (i.e. substituting $u_i = \sqrt{2/\beta_e} u_i'$, 
$\varphi = \sqrt{2/\beta_e} \varphi'$, $\partial_t= \sqrt{2/\beta_e} \partial_t'$)
and neglecting the electron inertia, we recover the reduced Hall-magnetohydrodynamics (RHMHD) equations (E19)
and (E20) of Ref. \cite{Schekochihin09}. Furthermore, 
as noted in  Ref. \cite{Boldyrev15}, when concentrating on Alfv\'en waves and 
thus neglecting the
coupling to $u_i$, one easily checks that, in the present low $\beta_e$
limit where the coefficient $1+ 2/\beta_e$
in  Eq. (\ref{eq:FP1}) reduces to $2/\beta_e$,
the $\beta_e$ parameter can be scaled out by writing  $B_z = \sqrt{\beta_e/2} B_z'$,
thus  making $\rho_s$ the only characteristic scale of this system.

When neglecting electron inertia, Eqs. (\ref{eq:FP1})-(\ref{eq:FP5})  can be
considered for any
value of $\beta_e$. In the large $\beta_e$ limit and in 2D, by using the above rescalings for $u_i$, $\varphi$ and time, the resulting system identifies with Eqs. (20)-(23) of 
Ref. \cite{Andres14b} for incompressible two-fluid MHD, when taking  $\varepsilon = \sqrt{2/\beta_e}$ which measures $d_i$ in units of $\rho_s$. Note that the system derived in Ref. \cite{Andres14b} involves an equation for $A_z$ instead of $A_\|$ 
(quantities which identify at the considered order). In this case, the last term of Eqs. (\ref{eq:FP2}) originates from the Hall term, while it here results from the electron pressure in Ohm's law. Pressure balance ensures the equality of these two contributions.



\subsection{The case of  small but finite ion temperature}

\subsubsection{Derivation of the ion FLR contributions}  \label{ssec:temp}
Since with scaling II, $\rho_i/\rho_s=O(\delta^{1/2})$, ion FLR corrections enter the dynamics as contributions of  order $\delta$.
Using this scaling, we first derive the electron equations. At the required order in Eq. (\ref{eq:Apar}), we have $\varphi_e=\varphi-n$ and, from Ref.\cite{PST17},
\begin{equation}
{\widehat{\boldsymbol b}} \bcdot  (\bnabla_\perp \bcdot {\boldsymbol \Pi}_e)= \delta^2 [n, u_e].
\end{equation}
This contribution  cancels the diamagnetic drift $\delta^2 [n, u_e]$ that originates from  the second term of $\varphi_e$.
Equation (\ref{eq:Apar}) thus rewrites
\begin{equation}
\frac{d}{dt} (A_\| -\delta^2 u_e) +\partial_z \varphi -\nabla_\| n=0.\label{eq:AparII}
\end{equation}

In Eq. (\ref{eq:Bz}), $B_z$ is negligible as well as the non-gyrotropic pressure contribution. This equation thus reduces to
\begin{equation}
\frac{d}{dt} n +\nabla_\| u_e =0.\label{eq:nII}
\end{equation}

We now turn to the velocity equations (\ref{eq:upari}) and (\ref{eq:vorticities}).
The  ion non-gyrotropic pressure tensor can  be estimated within a perturbative computation in terms of the parameters $\varepsilon$ and $\tau$
from the coupled system provided by Eq. (A6) of Ref. \cite{Scheko10} and a drift 
expansion of the ion transverse velocity. Neglecting the heat flux contributions
to ${\boldsymbol \Pi}_i$, we are led, in practice,  to repeat the calculations made 
in Appendix A of Ref. \cite{PST17}, only replacing
pressures and velocities of the electrons  by those of the ions and
dropping the factors $-\delta^2$  and $\delta^4$, which corresponds to changing the charge and the mass when replacing electrons by ions.
This results in expressing the 
parallel component of the nongyrotropic ion pressure force as 
\begin{eqnarray}
&&\bhatb \bcdot  (\bnabla_\perp \bcdot {\boldsymbol \Pi}_i)=
-[p_{\perp i} -B_z, u_i] - \nabla_\| \Delta_\perp \varphi_i \nonumber \\
&&\qquad - [\bnabla_\perp \varphi_i ; \bnabla_\perp A_\|]
  -\partial_t \Delta_\perp u_i - [\varphi_i, \Delta_\perp u_i] \nonumber \\
&& \qquad +\frac{1}{2} [\Delta_\perp \varphi_i, u_i] - [\bnabla_\perp \varphi_i; \bnabla_\perp u_i],
\end{eqnarray}
where we use the notation $\displaystyle{[\bnabla f;\bnabla g] = \sum_i [\partial_i f, \partial_i g]}$.
Equation (\ref{eq:upari}) then rewrites
\begin{eqnarray}
&&\partial_t (u_i - \tau \Delta_\perp u_i + \delta^2 u_e) + [\varphi_i, u_i - \tau 
\Delta_\perp u_i]  + [\varphi, \delta^2 u_e] \nonumber \\
&& \qquad - \tau [p_{\perp i}-\frac{1}{2}\Delta_\perp \varphi_i, u_i ]
- \tau [\bnabla_\perp \varphi_i;\bnabla_\perp (A_\|+u_i)]\nonumber \\
&& \qquad + \nabla_\| ( n + \tau p_{\| i}- \tau \Delta_\perp\varphi_i) =0. \label{eq:ui}
\end{eqnarray}
  For the vorticity equation, we need to express
\begin{eqnarray}
&&\bhatb\bcdot \bnabla_\perp \times (\bnabla_\perp \bcdot {\boldsymbol \Pi}_i)=
  -[p_{\perp i},\Delta_\perp \varphi_i]- [\bnabla_\perp p_{\perp i}; \bnabla_\perp \varphi_i]\nonumber \\
  && + \frac{1}{2}\nabla_\| \Delta_\perp u_i +\frac{1}{2}[\Delta_\perp A_\|, u_i]
  + \frac{1}{2} \Delta_\perp (\bnabla \bcdot {\boldsymbol u}_i)\nonumber \\
  && -\frac{1}{4} \left (\partial _t \Delta_\perp^2 \varphi_i  + [\varphi_i,\Delta_\perp^2 \varphi_i] \right )
  - [\bnabla_\perp \varphi_i; \bnabla_\perp \Delta_\perp \varphi_i], \label{brotdivPi}
 \end{eqnarray}
where the last line of Eq. (\ref{brotdivPi}) is obtained by a computation to
second order in terms of scale separation. The latter computation is rather cumbersome and was performed using MAPLE symbolic 
calculation software. In this expression, it is of interest to rewrite
\begin{eqnarray}
  &&\Delta_\perp (\bnabla \bcdot {\boldsymbol u}_i) = -\Delta_\perp (\partial_t n
  + [\varphi_i, n]) = -\partial_t \Delta_\perp n \nonumber \\
&&\quad - [\varphi_i, \Delta_\perp n] - [\Delta_\perp \varphi_i, n]
  - 2 [\bnabla_\perp \varphi_i; \bnabla_\perp n],
\end{eqnarray}
where  one can make the replacement
\begin{equation}
\partial_t \Delta_\perp n =
- \Delta_\perp \left ([\varphi_i + \frac{\tau}{2} \Delta_\perp ·\varphi_i,n]
+ \nabla_\| u_e \right), \label{lapldens}
\end{equation}
the second term in the bracket becoming subdominant when substituted into the vorticity equation.

At the considered order, noting that the contribution of the ion gyrotropic pressure is of lower order, the vorticity equation becomes, after writing $p_{\perp i}=n+t_{\perp i}$, where 
$t_{\perp i}$ refers to the  perpendicular ion temperature fluctuations (and  $t_{\| i}$ to the parallel ones),

\begin{eqnarray}
&& \partial_t \left (\Delta_\perp \varphi_i  - \frac{\tau}{4} \Delta_\perp^2 \varphi_i\right)
+ [\varphi_i, \Delta_\perp \varphi_i  -\frac{\tau}{4}\Delta_\perp^2 \varphi_i] 
\nonumber \\
&&+\frac{2}{\beta_e} \nabla_\|\Delta_\perp A_\| + \frac{\tau}{2} \Delta_\perp \nabla_\| u_e
+ \tau[\Delta_\perp \varphi_i, n] \nonumber \\
&&+ \tau[\bnabla_\perp \varphi_i; \bnabla_\perp (n - \Delta_\perp \varphi_i)]
-\tau \bnabla_\perp \bcdot [t_{\perp_i} , \bnabla_\perp \varphi_i]=0. \label{eq:phiII}\nonumber\\
\end{eqnarray}
\noindent
\textit {Determination of the temperature fluctuations:}
As discussed in Appendix \ref{app:dispersion}, the present scaling suggests  considering an adiabatic regime for the ions, where gyrotropic heat fluxes are negligible. In this case, neglecting also the fourth-rank cumulant contributions
($B_z$ being  small in the present ordering),
one has 
\begin{equation}
\frac{d^i}{dt}t_{\parallel_i} + 2 \nabla_\| u_i 
	+ \tau [ t_{\parallel_i} , p_{\perp i}]=0
\end{equation}
or
\begin{equation}
\frac{d}{dt} t_{\parallel_i} - \frac{\tau}{2} [\Delta_\perp \varphi,t_{\parallel_i}]
+ 2 \nabla_\| u_i  =0.
\end{equation}
Similarly,
\begin{equation}
\frac{d^{i}}{dt} (t_{\perp_i}  - n) - \nabla_\| u_i 
+ 2\tau [t_{\perp_i} , p_{\perp i}]=0,
\end{equation}
which rewrites
\begin{equation}
\frac{d}{dt}(t_{\perp_i} -n)
- \frac{\tau}{2} [\Delta_\perp \varphi, t_{\perp_i} -n] - \nabla_\| u_i  =0.
\end{equation}
The terms of the form $\nabla_\| u_i$ are subdominant within scaling II. If one is not interested in the own dynamics of the temperatures, they only need to be determined at the dominant order, and it is possible to take 
\begin{eqnarray}
&&\frac{d}{dt}t_{\parallel_i}=0 \label{eq:tpar}\\
&&\frac{d}{dt}t_{\perp_i}=\frac{d}{dt}n=-\nabla_\| u_e.\label{eq:tperp}
\end{eqnarray}
Since we also have
\begin{equation}
\Delta_\perp A_\| = \frac{\beta_e}{2} u_e, \label{eq:ueII}
\end{equation}
we conclude that, within scaling II, the equation for $u_i$ is decoupled.
The system of Eqs. (\ref{eq:phiII}), (\ref{eq:AparII}), (\ref{eq:nII}), together with (\ref{eq:ueII}), (\ref{eq:tperp}) and the relation $\varphi= \varphi_i + \frac{\tau}{2} \Delta_\perp \varphi_i - \tau n- \tau t_{\perp_i}$, conserves the energy
\begin{eqnarray}
{\mathcal E_1} &=& \frac{1}{2} \int \Big (|\bnabla_\perp \varphi_i|^2 +\delta^2 u_e^2 
+ \frac{\tau}{4}(\Delta_\perp \varphi_i)^2 \nonumber \\
&+& \frac{2}{\beta_e}|\bnabla_\perp A_\||^2
+(1+ \tau) n^2  + \tau t_{\perp_i}^2 \Big )
d^3{\boldsymbol x}.
\end{eqnarray}
A further simplification is possible (with a proper choice of initial conditions) where temperatures are determined algebraically. For this purpose, one can remark that
the number density $n$ is also given by the ion continuity equation in the form (after using the expression for $\varphi_i$)
\begin{equation}
\frac{dn}{dt} - \frac{\tau}{2} [\Delta_\perp \varphi, n] 
+ \tau [p_{\perp i}, n] + \bnabla \cdot {\boldsymbol u}_i =0.
\end{equation}
In order to estimate $\bnabla_\perp \cdot {\boldsymbol u}_{\perp i}$,
we consider the drift expansion  of the transverse velocity.
\begin{eqnarray}
{\boldsymbol u}_{\perp i} &=& \frac{1}{B}  \bhatb  \times  \Big \{
\bnabla_\perp \varphi + \frac{\tau}{1+n} \bnabla_\perp
p_{\perp i} + \frac{\tau}{1+n} (\bnabla_\perp
\bcdot {\boldsymbol \Pi}_i) \nonumber \\
&+& \partial_t {\boldsymbol A}_\perp  
+ \frac{d^{(i)}}{dt} {\boldsymbol u}_{\perp i} \Big \}. \label{drift}
\end{eqnarray}
where $B= |{\boldsymbol B}|= 1 + B_z + O(\varepsilon^2)$
and ${\boldsymbol A}_\perp$ is the transverse component of magnetic vector potential.
As  at scales comparable to $\rho_s$, $B_z = O(\varepsilon \beta_e)$ and 
${\boldsymbol A}_\perp$  also scales as $\varepsilon \beta_e$, it follows that
\begin{equation}
\bhatb \times \frac{d^{(i)}}{dt} {\boldsymbol u}_{\perp i} = -\partial_t \bnabla \varphi- [\varphi, \bnabla_\perp \varphi] 
+ O(\varepsilon^2\tau).
\end{equation}
As one also has $\bnabla_\perp 	\bcdot (\bhatb  \times \bnabla_\perp \varphi) =
O(\varepsilon^2 (\varepsilon + \beta_e ))$,
it follows that
\begin{equation}
\bnabla_\perp 	\bcdot {\boldsymbol u}_{\perp i} = 
- \frac{d}{dt}\Delta_\perp \varphi + O(\varepsilon^2 (\tau + 
\varepsilon)), 
\end{equation}
and consequently
\begin{equation}
\frac{d}{dt}(t_{\perp_i}- \Delta_\perp \varphi)  = 
O(\varepsilon^2(\tau +  \varepsilon)). \label{T-lapphi}
\end{equation}
For suitable initial conditions, one can thus write $t_{\| i} =0$ and $t_{\perp_i} = \Delta_\perp \varphi$,
which reproduces the closure for the perpendicular ion temperature used in
Ref. \cite{Hsu86}.
The system can then be reduced to a 3-field model
made up of Eqs. (\ref{eq:AparII}), (\ref{eq:nII}) and of the equation for the parallel vorticity
\begin{eqnarray}
&& \partial_t \left (\Delta_\perp \varphi_i  - \frac{\tau}{4} \Delta_\perp^2 \varphi_i\right)
+ [\varphi_i, \Delta_\perp \varphi_i  -\frac{\tau}{4}\Delta_\perp^2 \varphi_i] 
\nonumber \\
&&\qquad + \frac{2}{\beta_e} \nabla_\|\Delta_\perp A_\|+ \frac{\tau}{2} \Delta_\perp \nabla_\| u_e
+ \tau[\Delta_\perp \varphi_i, n] \nonumber \\
&&\qquad + \tau[\bnabla_\perp \varphi_i; \bnabla_\perp n ]=0, \label{eq:phiIIbis}
\end{eqnarray}
together with Eq. (\ref{eq:ueII}) and the expression for $\varphi$ in terms of $\varphi_i$, which now rewrites 
\begin{equation}   \label{varphii}
\varphi= \varphi_i - \frac{\tau}{2} \Delta_\perp \varphi_i - \tau n.
\end{equation}
This system does not conserve energy. In a way similar to what is done in Ref. \cite{Hsu86}, 
adding to Eq. (\ref{eq:phiIIbis}) the equation
\begin{equation}
-\tau(\partial_t \Delta_\perp^2\varphi_i+[\varphi_i,\Delta_\perp^2\varphi_i]+\Delta_\perp\nabla_\| u_e +2[\nabla\varphi_i;\nabla\Delta_\perp\varphi_i])=0,
\end{equation}
obtained after taking the Laplacian of the vorticity equation at dominant order, we obtain
a new system, equivalent to the previous one at order $O(\tau)$
in the form
\begin{eqnarray}
&&\partial_t \left (\Delta_\perp \varphi^*  - \frac{5}{4}{\tau} \Delta_\perp^2 \varphi^*\right)
+ [\varphi^*, \Delta_\perp \varphi^*  -\frac{5}{4}{\tau} \Delta_\perp^2 \varphi^*] 
\nonumber \\
&&\qquad +\frac{2}{\beta_e} \nabla_\|\Delta_\perp A_\| - \frac{\tau}{2} \Delta_\perp \nabla_\| u_e
+ \tau[\Delta_\perp \varphi^*, n] \nonumber\\
&& \qquad + \tau[\bnabla_\perp \varphi^*; \bnabla_\perp (n-2\Delta_\perp\varphi^*)]=0 \\
&& \frac{d}{dt} n + \nabla_\| u_e = 0 \label{ne2f}\\
&&\frac{d}{dt} (A_\| -\delta^2 u_e)  + \partial_z \varphi - \nabla_\| n = 0, \label{ue2f}
\end{eqnarray}
where we introduced a new potential
\begin{equation}
\varphi^*=\varphi+\tau n +(\tau/2)\Delta_\perp \varphi^*,
\end{equation}
The above system conserves the energy
\begin{eqnarray}
{\mathcal E_2} &=& \frac{1}{2} \int \Big (|\bnabla_\perp \varphi^*|^2 +\delta^2 u_e^2 
+ \frac{5\tau}{4}(\Delta_\perp \varphi^*)^2 + \frac{2}{\beta_e}|\bnabla_\perp A_\||^2\nonumber \\
&&+(1+ \tau) n^2 \Big )
d^3{\boldsymbol x}.
\end{eqnarray}

This model introduces ion FLR corrections but neglects the coupling with the ion parallel velocity. The ordering is indeed limited to scales where $u_e$ is much larger than $u_i$, a condition which excludes scales of order  $d_i$ or larger. 

\subsubsection{Extension of the model to larger scales}  \label{ssec:templs}

At larger scales, another scaling (scaling III) must be used where, keeping $\beta_e=O(\delta^2)$ and $\tau=O(\delta)$, one assumes $\bnabla_\perp \sim \delta$, $\varphi = O(\varepsilon)$,
$n \sim u_i \sim u_e \sim A_\| = O(\delta\varepsilon)$,
$\partial_t \sim \delta^2 \varepsilon$ and
$\partial_z \sim \delta^3 \varepsilon$. In this regime, the system takes the form of the RHMHD equations (in the small $\beta$ limit), where electron inertia and finite Larmor radius corrections are absent. It is then easy to build a uniform model that reduces to the latter large-scale model or to the former 3-field model when scalings III or II are applied respectively. It contains terms that are negligible in one or the other specific limits, and also sub-dominant additional  terms, corresponding to the first two terms of the second line of 
Eq. (\ref{brotdivPi}), needed  for the energy to be conserved.

Keeping the dynamical equations for the temperature fluctuations but neglecting the $O(\tau)$ corrections which turn out to be irrelevant at the order of the
asymptotics, we are led to write the reduced fluid model in the form
\begin{eqnarray}
&& \partial_t \left (\Delta_\perp \varphi_i  - \frac{\tau}{4} \Delta_\perp^2 \varphi_i\right)
+ [\varphi_i, \Delta_\perp \varphi_i  -\frac{\tau}{4}\Delta_\perp^2 \varphi_i] 
\nonumber \\
&&\qquad +\frac{2}{\beta_e} \nabla_\|\Delta_\perp A_\| + \frac{\tau}{2} \Delta_\perp \nabla_\| u_e
+ \tau[\Delta_\perp \varphi_i, n] \nonumber \\
&& \qquad+ \tau[\bnabla_\perp \varphi_i; \bnabla_\perp (n - \Delta_\perp \varphi_i)]
+ \frac{\tau}{2}\nabla_\| \Delta_\perp u_i  \nonumber \\
&& \qquad+\frac{\tau}{2}[\Delta_\perp
A_\|, u_i] -\tau \bnabla_\perp \bcdot [t_{\perp_i} , \bnabla_\perp \varphi_i]=0 \label{lapphi}\\
&&\partial_t (u_i - \tau \Delta_\perp u_i + \delta^2 u_e) + [\varphi_i, u_i - \tau 
\Delta_\perp u_i]  + [\varphi, \delta^2 u_e] \nonumber \\
&& \qquad - \tau [p_{\perp i}-\frac{1}{2}\Delta_\perp \varphi_i, u_i ]
 - \tau [\bnabla_\perp \varphi_i;\bnabla_\perp (A_\|+u_i)]\nonumber \\
 && \qquad + \nabla_\| ( n + \tau p_{\| i}- \tau \Delta_\perp\varphi_i) =0 \label{eq:uils}\\
&& \frac{d}{dt} n  + \nabla_\| u_e = 0 \label{density}\\
&&\frac{d}{dt} (A_\| -\delta^2 u_e) + 
\partial_z \varphi - \nabla_\| n = 0 \\
&&\frac{d}{dt} t_{\parallel_i} + 2 \nabla_\| u_i  =0  \label{tpi}\\
&&\frac{d}{dt}(t_{\perp_i} -n) - \nabla_\| u_i  =0  \label{tpei}\\
&&\Delta_\perp A_\| = \frac{\beta_e}{2}(u_e - u_i)\\
&&\varphi = \varphi_i + \frac{\tau}{2} \Delta_\perp\varphi_i - \tau p_{\perp i}\\
&& p_{\perp i} = n + t_{\perp_i} \quad , \quad p_{\| i} = n +t_{\parallel_i}.
\end{eqnarray}
The energy is given by
\begin{eqnarray}
{\mathcal E}_3 &=& \frac{1}{2} \int \Big (u_i^2 + \tau |\bnabla_\perp u_i|^2
+ |\bnabla_\perp \varphi_i|^2 +\delta^2 u_e^2 
+ \frac{\tau}{4}(\Delta_\perp \varphi_i)^2 \nonumber \\
&+& \frac{2}{\beta_e}|\bnabla_\perp A_\||^2
+(1+ \tau) n^2  + \tau t_{\perp_i}^2 + \frac{\tau}{2}{t_{\parallel_i}}^2 \Big )
d^3{\boldsymbol x}.
\end{eqnarray}

Similarly to what was done at the level of the 3-field model, it is possible to simplify this system (assuming suitable initial conditions) by prescribing $t_{\parallel_i}=0$ and $t_{\perp_i}=\Delta_\perp \varphi$ (or equivalently, at the level of the present ordering, $t_{\perp_i}=\Delta_\perp \varphi^*$) and perform the same combination with the Laplacian of the vorticity equation in order to ensure energy conservation. In this case, we obtain
\begin{eqnarray}
&& \partial_t \left (\Delta_\perp \varphi^*  - \frac{5\tau}{4} \Delta_\perp^2 \varphi^*\right)
+ [\varphi^*, \Delta_\perp \varphi^*  -\frac{5\tau}{4}\Delta_\perp^2 \varphi^*] 
\nonumber \\
&&\qquad +\frac{2}{\beta_e} \nabla_\|\Delta_\perp A_\| - \frac{\tau}{2} \Delta_\perp \nabla_\| u_e
+ \tau[\Delta_\perp \varphi^*, n] \nonumber \\
&& \qquad+ \tau[\bnabla_\perp \varphi^*; \bnabla_\perp (n-2\Delta_\perp\varphi^*) ]
+ \frac{\tau}{2}\nabla_\| \Delta_\perp u_i \nonumber\\
&&\qquad+\frac{\tau}{2}[\Delta_\perp
A_\|, u_i] =0 \label{eq:uniform_lapphi}\\
&&\partial_t (u_i - \tau \Delta_\perp u_i + \delta^2 u_e) + [\varphi^*, u_i - \tau 
\Delta_\perp u_i]  + [\varphi, \delta^2 u_e] \nonumber \\
&& \qquad - \tau [n+\frac{1}{2}\Delta_\perp \varphi^*, u_i ]
- \tau [\bnabla_\perp \varphi^*;\bnabla_\perp (A_\|+u_i)]\nonumber \\
&& \qquad + \nabla_\| ((1+\tau) n - \tau \Delta_\perp\varphi^*) =0 \label{eq:uniform_ui}\\
&& \frac{d}{dt} n  + \nabla_\| u_e = 0 \label{eq:uniform_density}\\
&&\frac{d}{dt} (A_\| -\delta^2 u_e) + 
\partial_z \varphi - \nabla_\| n = 0 \label{eq:uniform_Apar}\\
&&\Delta_\perp A_\| = \frac{\beta_e}{2}(u_e - u_i)\\
&&\varphi=\varphi^* -(\tau/2)\Delta_\perp \varphi^*-\tau n,  \label{varphistar}
\end{eqnarray}
which provides a four-field model valid from the MHD to the sub-$d_e$ scales, in the regime where the parameters $\beta_e$ and $\tau$ are both small.

  For this system, the energy reads
\begin{eqnarray}
{\mathcal E}_4 &=& \frac{1}{2} \int \Big (u_i^2 + \tau |\bnabla_\perp u_i|^2
+ |\bnabla_\perp \varphi_i|^2 +\delta^2 u_e^2 
+ \frac{5\tau}{4}(\Delta_\perp \varphi_i)^2 \nonumber \\
&+& \frac{2}{\beta_e}|\bnabla_\perp A_\||^2
+(1+ \tau) n^2   \Big ) d^3{\boldsymbol x},
\end{eqnarray}

When taking $\tau=0$ and recalling that $n=-(2/\beta_e)B_z$, this system reduces to Eqs. (\ref{eq:FP1})-(\ref{eq:FP5}) where, in Eq. (\ref{eq:FP1}), the coefficient 1 is neglected compared to  $2/\beta_e$.

\section{Gyrofluid modeling for arbitrary $\tau$}  \label{sec:gyro}

In this Section,  we consider as the starting point the gyrofluid system (\ref{neg})-(\ref{amppeg}) 
which allows considering  all the values of the ion-electron temperature ratio.
As a first step, it is of interest to reproduce 
the reduced fluid models of Secs. \ref{sec:notemp}, \ref{ssec:temp} and \ref{ssec:templs}, using the corresponding scalings with regard to particle moments, electromagnetic fields, parameters, length and time scales. In addition, we specify orderings for the gyrofluid moments.
This comparison is of interest in that it points out 
that consistency between the two approaches requires the prescription of  closure 
relations that are consistent with the assumed scalings. In this context, we recall that previous analyses of relations between gyrofluid and FLR reduced fluid models were carried out in Refs. \cite{Bri92,Sco07,Bel01}.

 In all three cases, it is understood that the electron fluid is assumed to be isothermal and that contributions due to heat flux and energy-weighted pressure tensors in the ion fluid equations are negligible. Also, we assume negligible {\it gyrofluid} ion perpendicular temperature fluctuations, i.e. $\Ppei-N_i=0$. Denoting by $T_{\perp_\alpha}=P_{\perp_\alpha}-N_{\alpha}$ and $T_{\parallel_\alpha}=P_{\parallel_\alpha}-N_\alpha$ the perpendicular and parallel gyrofluid temperature fluctuations related to the species $\alpha$, we remark that the assumption $T_{\perp_i}=0$ is satisfied if the underlying perturbation of the ion gyrocenter distribution function $\tilde{F}_i$, in dimensional form, is given by
\beq
\widetilde{F}_i=F_{eq\,i}\left(\frac{\widetilde{N}_i}{n_0}+2\frac{v}{v_{th\,i}}\frac{\widetilde{U}_i}{v_{th\,i}}+\frac{1}{2}\left(2\frac{v^2}{v_{th \,i}^2} -1\right)\frac{\widetilde{T}_{\parallel i}}{T_{0i}}\right)
\eeq
where the tilde denotes a dimensional quantity, $v_{th \,i}=\sqrt{2\tau}c_s$ is the thermal ion speed and 
\beq  \label{maxw}
F_{eq \,i}(v,\mu)=n_0\left({\frac{m_i}{2\pi T_{0i}}}\right)^{3/2}\exp\left(-m_i \frac{v^2}{2 T_{0i}}-\frac{\mu B_0}{T_{0i}}\right),
\eeq
is an equilibrium Maxwellian distribution function with $v$ and $\mu$ indicating the parallel velocity and the ion magnetic moment, respectively. We remark that this choice of 
$F_{eq \,i}$ yields  $\Qpei=\Qpi=R_{\parallel \perp_{i}}=R_{\perp \perp_{i}}=0$, which is consistent with the above assumption of neglecting heat flux and energy-weighted pressure tensor contributions.

 Finally, Alfv\'en speed is assumed to be non-relativistic, i.e. $v_A \ll c$. 
\subsection{Small ion temperatures}

\subsubsection{Negligible ion temperature}   \label{ssec:notempg}

In order to derive a cold-ion model, we assume 
\begin{align}
&\beta_e=O(\delta), \qquad \tau=O(\delta^{3/2}), \qquad \nabla_\perp=O(1),  \label{ord11}\\
&U_e\sim u_e=O\left(\frac{\varepsilon}{\delta^{1/2}}\right), \qquad \bpar=O(\delta \varepsilon),\\
&\apar \sim \partial_z \sim U_i \sim u_i= O(\delta^{1/2} \varepsilon),\\
&\partial_t \sim N_{e,i}  \sim \varphi \sim P_{\parallel_{e,i}} \sim P_{\perp_{e,i}}  \nno \\
&\sim n_{e,i} \sim p_{\parallel_{e,i}} \sim p_{\perp_{e,i}}=O(\varepsilon).  \label{ord14}
\end{align}
Ordering (\ref{ord11})-(\ref{ord14}), devoid of gyrofluid variables, corresponds to scaling I of Sec. \ref{sec:notemp}.

We apply  ordering   (\ref{ord11})-(\ref{ord14}), together with the above assumptions on the closures and the non-relativistic character of the Alfv\'en speed, to Eqs. (\ref{neg}), (\ref{ueg}), (\ref{nig}), (\ref{uig}), (\ref{poissg}), (\ref{amppg}), (\ref{amppeg}). Retaining, in each dynamical equation, the leading order terms and the corrections of order $\delta$, we obtain
\begin{align}
&\frac{\partial N_e}{\pa t}+[\varphi, N_e]-[\bpar , \Ppee]+\gpar U_e=0,  \label{neg1}\\
&\frac{\pa}{\pa t}( \delta^2 U_e - \apar)+[\varphi , \delta^2 U_e - \apar] \nno \\
&+\gpar(\Ppe + \bpar)-\pa_z\varphi=0, \label{ueg1}\\
&\frac{\pa N_i}{\pa t}+[\varphi , N_i]+\gpar U_i=0, \label{nig1}\\
&\frac{\pa}{\pa t}(U_i + \apar)+[\varphi , U_i + \apar]+\pa_z \varphi=0, \label{uig1}\\
&0=N_e-N_i-\lapp \varphi, \label{qng1}\\
&\lapp \apar = \frac{\beta_e}{2}(U_e - U_i), \label{amppg1}\\
&\bpar=-\frac{\beta_e}{2}(\Ppee + 2 \bpar). \label{amppeg1}
\end{align}
The evolution equations for $\Ppe$, $\Ppee$ and $\Ppei$ have not been considered because the closure relations will replace them. The evolution equation for $\Ppi$ is not necessary either, because the ordering made the contribution of $\Ppi$ in Eq. (\ref{uig1}) negligible, thus decoupling the evolution of the ion gyrofluid parallel pressure. 

In order to express the system (\ref{neg1})-(\ref{amppeg1}), closed with the electron isothermal relation $\ppe=\ppee=n_e$, in terms of particle moments, it is necessary to resort to the transformation from gyrofluid to particle moments \cite{Bri92} which, for the scaling under consideration, accounting for corrections of order $\delta$, reads
\begin{align}
&N_e=n_e - \bpar, \qquad U_e=u_e,  \label{tr11}\\
&\Ppe=\ppe-\bpar, \qquad \Ppee=\ppee-2\bpar,\\
&N_i=n_i-\lapp\varphi-\bpar, \qquad U_i=u_i,\\
&\Ppi=\ppi-\lapp\varphi-\bpar, \qquad \Ppei=\ppei-2\lapp\varphi-2\bpar.  \label{tr14}
\end{align}
%

Making use of the aforementioned electron isothermal closure, after inserting  relations (\ref{tr11})-(\ref{tr14}) into Eqs. (\ref{qng1})-(\ref{amppeg1}), we get
\begin{align}
&n_e=n_i=n, \label{qn1}\\
&\lapp \apar=\frac{\beta_e}{2}(u_e-u_i),  \label{ampp1}\\
&\bpar=-\frac{\beta_e}{2}\ppee=-\frac{\beta_e}{2}n.  \label{amppe1}
\end{align}
Inserting the transformations (\ref{tr11})-(\ref{tr14}) into Eqs. (\ref{neg1})-(\ref{uig1}), retaining only first order corrections in $\delta$, and making use of relations (\ref{qn1}) and (\ref{amppe1}), we obtain the system
\begin{align}
&\frac{d}{dt}(1+\frac{2}{\beta_e})B_z-\nabla_\| u_e=0\label{eq:FP1g}\\
&\frac{d}{dt}(A_\| -\delta^2 u_e)+\partial_z\varphi+\frac{2}{\beta_e}\nabla_\| B_z=0\label{eq:FP2g}\\
& \frac{d}{dt}(u_i+\delta^2 u_e)-\frac{2}{\beta_e}\nabla_\| B_z=0\label{eq:FP3g}\\
&\frac{d}{dt} \Delta_\perp\varphi+\frac{2}{\beta_e}\nabla_\|\Delta_\perp A_\|=0 \label{eq:FP4g},
\end{align}
which, together with Eq. (\ref{ampp1}), coincides with the system (\ref{eq:FP1})-(\ref{eq:FP5}) derived from a two-fluid description.

\subsubsection{Derivation of the ion FLR contributions}  \label{ssec:flr}

We consider here the ordering
\begin{align}
&\beta_e=O(\delta^2 ), \qquad \tau=O(\delta), \qquad \gperp=O(1),  \label{ord21}\\
&U_e\sim u_e =O\left(\frac{\varepsilon}{\delta}\right), \qquad \bpar=O(\delta^2 \varepsilon),\\
&\apar \sim \pa_z \sim U_i \sim u_i=O(\delta \varepsilon),\\
&\partial_t \sim N_{e,i} \sim \varphi \sim P_{\parallel_{e,i}} \sim P_{\perp_{e,i}} \nno \\
&\sim n_{e,i} \sim p_{\parallel_{e,i}} \sim p_{\perp_{e,i}}=O(\varepsilon),  \label{ord24}
\end{align}
which corresponds to the scaling II treated in Sec. \ref{ssec:temp}.

Applying  ordering (\ref{ord21})-(\ref{ord24}) to Eqs.  (\ref{neg}), (\ref{ueg}), (\ref{nig}), (\ref{uig}), (\ref{pig}), (\ref{poissg}), (\ref{amppg}), imposing $\Ppei-N_i=0$, neglecting the term proportional to $v_A^2/c^2$ in Eq. (\ref{poissg}) and retaining leading order terms as well as corrections of order $\tau$ (or, equivalently, of order $\delta$), we obtain
\begin{align}
&\frac{\partial N_e}{\pa t}+[\varphi, N_e]+\gpar U_e=0,  \label{neg2}\\
&\frac{\pa}{\pa t}( \delta^2 U_e - \apar)+[\varphi , \delta^2 U_e - \apar] +\gpar\Ppe -\pa_z \varphi=0, \label{ueg2}\\
&\frac{\pa N_i}{\pa t}+[\varphi, N_i]+\frac{\tau}{2}[\lapp \varphi , N_i]=0, \label{nig2}\\
&\frac{\pa}{\pa t}\left(U_i + \apar + \frac{\tau}{2}\lapp \apar\right)+[\varphi,
U_i + \apar + \frac{\tau}{2}\lapp \apar] \nno\\
&+\frac{\tau}{2}[\lapp \varphi , U_i]+\gpar\left(\tau \Ppi +\frac{\tau}{2}\lapp \varphi\right)+\pa_z \varphi=0,  \label{uig2}\\
&\frac{\pa \Ppi}{\pa t}+[\varphi , \Ppi]+\frac{\tau}{2}[\lapp\varphi , \Ppi]=0, \label{pig2}\\
&0=N_e-N_i-\lapp \varphi-\frac{\tau}{2}\lapp N_i-\frac{3}{4}\tau\bilapp\varphi, \label{poissg2}\\
&\lapp \apar=\frac{\beta_e}{2}U_e.  \label{ampg2}
\end{align}
Unlike the case of  ordering (\ref{ord11})-(\ref{ord14}),  with  ordering (\ref{ord21})-(\ref{ord24}), parallel magnetic perturbations become negligible, so that we did not invoke Eq. (\ref{amppeg}). Also, parallel ion gyrofluid velocity contributions become subdominant in Eq. (\ref{nig2}). Nevertheless, we determined also Eqs. (\ref{uig2}) and (\ref{pig2}) which, although decoupled with the present scaling, become relevant in the extended model accounting also for larger scales.

Based on scaling  (\ref{ord21})-(\ref{ord24}), the transformation from gyrofluid to particle moments becomes
\begin{align}
&N_e=n_e, \qquad U_e=u_e,  \label{tr21}\\
&\Ppe=\ppe, \qquad \Ppee=\ppee,\\
&N_i=n_i-\lapp\varphi-\frac{\tau}{2}\lapp n_i-\frac{\tau}{4}\bilapp\varphi, \label{tr23} \\
& U_i=u_i-\frac{\tau}{2}\lapp u_i,  \label{tr24}\\
&\Ppi=\ppi-\lapp\varphi-\frac{\tau}{2}\lapp \ppi-\frac{\tau}{4}\bilapp\varphi, \label{tr25} \\
&\Ppei=\ppei-2 \lapp\varphi-\tau \lapp\ppei-\frac{9}{4}\tau \bilapp \varphi.  \label{tr26}
\end{align}
Applying this transformation to Eqs. (\ref{neg2}), (\ref{ueg2}), (\ref{nig2}), (\ref{poissg2}) and (\ref{ampg2}), retaining first order corrections in $\tau$ and using  the assumptions on the closures for the electron fluid, we obtain after some algebra
\begin{align}
&\frac{d n}{dt}+\gpar u_e=0,  \label{ne2}\\
&\frac{d}{dt}(\apar - \delta^2 u_e)+\pa_z \varphi-\gpar n=0,  \label{ue2}\\
&\frac{d}{dt}\left(\lapp\varphi+\frac{\tau}{4}\bilapp\varphi\right)-\tau[\lapp\varphi ,n] \nno\\
&-\tau [\gperp\varphi ; \gperp n]+\gpar u_e-\frac{\tau}{2}\lapp\gpar u_e=0, \label{vort2}\\
&n_e=n_i=n,\\
&\lapp\apar =\frac{\beta_e}{2}u_e.  \label{amp2}
\end{align} 
We now remark that the continuity equation (\ref{ne2}) and the generalized Ohm's law (\ref{ue2}) correspond to Eqs. (\ref{eq:nII}) and (\ref{eq:AparII}), respectively. Combining Eq. (\ref{ne2}) with Eq. (\ref{vort2}) and introducing the potential $\varphi_i$ defined in Eq. (\ref{varphii}), we obtain Eq. (\ref{eq:phiIIbis}). Closing the system by means of  relation (\ref{amp2}), we then retrieve the 3-field model derived in Sec. \ref{ssec:temp}. With regard to ion temperature fluctuations,  due to the assumption $\Tpei=\Ppei-N_i=0$, from Eqs. (\ref{tr23}) and (\ref{tr26}), we obtain $\tpei=\lapp\varphi$, up to corrections of order $O(\tau\varepsilon)$, which is namely the hypothesis underlying the closure of the 3-field model, as derived from the two-fluid description. With regard to the parallel temperature, from Eqs. (\ref{nig2}) and (\ref{pig2}), after transforming into particle moments by means of Eqs. (\ref{tr23}) and (\ref{tr25}), we obtain, to leading order.
\beq
\frac{d \tpi}{dt}=0,
\eeq
which coincides with Eq. (\ref{eq:tpar}).

Finally, the decoupled parallel ion velocity equation, obtained from Eq. (\ref{uig2}) after transforming to particle moments, reads
\begin{align}
&\frac{d}{dt}(u_i-\tau\lapp u_i + \delta^2 u_e)+\gpar(n+\tau \ppi-\tau \lapp \varphi) \nno \\
&-\tau [\gperp\varphi ; \gperp(u_i+\apar)]=0. \label{ui2}
\end{align}
Equation (\ref{ui2}), after replacing $\varphi$ in favor of $\varphi_i$, coincides with Eq. (\ref{eq:ui}),  once   the above mentioned closure condition $\tpei=\lapp\varphi$ has been inserted in this equation,

\subsubsection{Extension to larger scales}  \label{ssec:templsg}

Analogously to Sec. \ref{ssec:templs}, we here consider  a scaling valid for scales much larger than $\rho_s$,  which introduces a coupling with the parallel ion velocity. The scaling reads
\begin{align}
&\beta_e=O(\delta^2 ), \qquad \tau\sim\gperp=O(\delta),  \label{ord31}\\
&N_{e,i}\sim U_{e,i}\sim P_{\parallel_{e,i}} \sim P_{\perp_{e,i}} \sim \apar \nno\\
&\sim n_{e,i} \sim u_{e,i}\sim p_{\parallel_{e,i}}  \sim p_{\perp_{e,i}}=O(\delta \varepsilon),\\
&\varphi=O(\varepsilon), \qquad \pa_t =O(\delta^2 \varepsilon),\\
&\pa_z \sim \bpar=O(\delta^3 \varepsilon),  \label{ord34}
\end{align}
and corresponds to scaling III.

Proceeding similarly to Secs. \ref{ssec:notempg} and \ref{ssec:flr}, from the parent gyrofluid model (\ref{neg})-(\ref{amppeg}), retaining first order corrections in $\delta$, we obtain, from  scaling (\ref{ord31})-(\ref{ord34}), the following equations
\begin{align}
 &\frac{\partial N_e}{\pa t}+[\varphi, N_e]+\gpar U_e=0,  \label{neg3}\\
&\frac{\pa \apar}{\pa t}+[\varphi ,  \apar] -\gpar\Ppe +\pa_z \varphi=0, \label{ueg3}\\
&\frac{\pa N_i}{\pa t}+[\varphi, N_i]+\gpar U_i=0, \label{nig3}\\
&\frac{\pa}{\pa t}\left(U_i + \apar \right)+[\varphi,
U_i + \apar ]+\pa_z \varphi=0,  \label{uig3}\\
&\frac{\pa \Ppi}{\pa t}+[\varphi , \Ppi]+3\gpar U_i=0, \label{pig3}\\
&0=N_e-N_i-\lapp \varphi\label{poissg3}\\
&\lapp \apar=\frac{\beta_e}{2}(U_e-U_i).  \label{ampg3}
\end{align}
As in the case of ordering (\ref{ord21})-(\ref{ord24}), parallel magnetic fluctuations become negligible. 

The transformation from gyrofluid to particle moments is in  this case  given by
\begin{align}
&N_e=n_e, \qquad U_e=u_e,  \label{tr31}\\
&\Ppe=\ppe, \qquad \Ppee=\ppee,\\
&N_i=n_i-\lapp\varphi, \qquad U_i=u_i\label{tr33} \\
&\Ppi=\ppi-\lapp\varphi, \qquad \Ppei=\ppei-2 \lapp\varphi. \label{tr34} 
\end{align}
Applying this transformation to Eqs. (\ref{neg3})-(\ref{ampg3}) yields, upon retaining first order corrections in $\delta$ and carrying out a few algebraic manipulations, the following equations
\begin{align}
&\frac{d n}{dt}+\gpar u_e=0,  \label{ne3}\\
&\frac{d \apar}{dt} +\pa_z \varphi-\gpar n=0,  \label{ue3}\\
&\frac{d \lapp\varphi}{dt}+\frac{2}{\beta_e}\gpar\lapp\apar=0, \label{vort3}\\
&\frac{d u_i}{dt}+\gpar n=0,  \label{ui3}\\
&\frac{d \tpi}{dt}+2\gpar u_i=0,  \label{tpi3}\\
&\frac{d}{dt}(\tpei-n)-\gpar u_i=0,  \label{tpei3}\\
&n_e=n_i=n,\\
&\lapp\apar =\frac{\beta_e}{2}(u_e-u_i). \label{amp3}
\end{align} 
The system composed by Eqs. (\ref{ne3}), (\ref{ue3}), (\ref{vort3}), (\ref{ui3}), (\ref{amp3}) corresponds to the RHMHD system in the small $\beta_e$ limit, which was the result of applying scaling III from the two-fluid approach, as mentioned in Sec. \ref{ssec:templs}. We added to such system the resulting evolution equations for the ion temperatures, corresponding to Eqs. (\ref{tpi3}) and (\ref{tpei3}). Equation  (\ref{tpi3}), expressed in terms of particle moments, descends from Eqs. (\ref{pig3}) and (\ref{nig3}),  whereas Eq. (\ref{tpei3}) can be obtained from Eq. (\ref{nig3}), when applying transformation (\ref{tr33}) and imposing, as previously assumed, $\lapp \varphi=\tpei$. Equations (\ref{tpi3}) and (\ref{tpei3}) coincide with Eqs. (\ref{tpi}) and (\ref{tpei}) respectively. 

We thus derived, in Secs. \ref{ssec:flr} and \ref{ssec:templsg}, by means of a gyrofluid approach, the same models derived from the two-fluid description using scalings II and III and imposing $\tpei=\lapp\varphi$ at leading order. The uniform model  (\ref{eq:uniform_lapphi})-(\ref{varphistar}) then directly follows
by applying the procedure adopted in Sec. \ref{ssec:templs}.

We remark that, although the model of Ref. \cite{Bri92} was taken as starting point for the gyrofluid derivation, for the model involving ion FLR corrections, other low-$\beta_e$ gyrofluid models, such as those of Refs. \cite{Sco10,Sny01}, could have been taken as parent models and would have led to the same result. The models of Refs. \cite{Sco10,Sny01} adopt different closures for the gyroaveraging operators, compared to Ref. \cite{Bri92}. However, as far as the first order corrections in $\tau$ are concerned, which is sufficient for our derivations, the different gyroaveraging operators yield the same expansion. On the other hand, the gyrofluid model of Ref. \cite{Bri92} accounts for parallel magnetic perturbations, which allows for the derivation of the model of Sec. \ref{ssec:notempg}, which refers to a higher $\beta_e$ regime.

\subsection{A two-field gyrofluid model for KAW dynamics}  \label{ssec:2fgyro}

The gyrofluid model presented in Appendix \ref{app:gyrofluid} greatly simplifies  when restricting to the evolution of the electron gyrocenter density and parallel velocity (assuming $N_i=T_{\perp i}=U_i=0$, with furthermore an isothermal assumption for the electrons, i.e. $T_{\| e}=0$ and $T_{\perp e}=-B_z$ as deduced from Eqs. (3.68a)-(3.69b) of Ref. \cite{TSP16}). Such a reduced system allows one to focus on Alfv\'en wave dynamics, neglecting the coupling with slow magnetosonic waves. It retains corrections associated with electron inertia and with temperature ratios of order up to $1/\beta_e \sim 1 /\delta$, which will in turn imply accounting also for an electron FLR contribution.
In order to derive the simplified gyrofluid model, we introduce two further scalings, denoted as scaling IV and V, respectively. Scaling IV is given by
\begin{align}
&\beta_e=O(\delta), \qquad \tau \sim \nabla_\perp=O(1),  \label{ord1a1}\\
&U_e = O\left(\frac{\varepsilon}{\delta^{1/2}}\right), \qquad \bpar=O(\delta \varepsilon),\\
&\apar \sim \partial_z = O(\delta^{1/2} \varepsilon),\\
&\partial_t \sim N_{e}  \sim \varphi =O(\varepsilon),  \label{ord1a4}
\end{align}
whereas scaling V corresponds to

\begin{align}
&\beta_e=O(\delta), \qquad \tau =O(1/\delta), \qquad  \nabla_\perp=O(1),  \label{ord1b1}\\
&U_e = O\left(\frac{\varepsilon}{\delta^{1/2}}\right), \qquad N_e \sim \bpar=O(\delta \varepsilon),\\
&\apar \sim \partial_z = O(\delta^{1/2} \varepsilon),\\
&\partial_t   \sim \varphi =O(\varepsilon).  \label{ord1b4}
\end{align}
Scaling V accounts for corrections relevant for large $\tau$ but is valid for smaller electron gyrocenter density fluctuations.

One then proceeds with applying the scalings IV and V to Eqs. (\ref{neg}), (\ref{ueg}), (\ref{poissg}), (\ref{amppg}) and (\ref{amppeg}), retaining the leading order terms, the corrections of order $\delta$ as well as one correction of order $\delta^2$ in Eq. (\ref{ueg}) which, as will be seen a posteriori, allows the final system to be cast in Hamiltonian form. Taking into account the closure relations mentioned at the beginning of Sec. \ref{ssec:2fgyro} and neglecting 
heat fluxes, as mentioned at the beginning of Sec. \ref{sec:gyro}, one obtains two closed systems. Retaining all terms present in both models, similarly with what was done in the case of the uniform model of Sec. \ref{ssec:templs}, one is led to the following two-field gyrofluid model
\begin{eqnarray}
&&\partial_t N_e +[\varphi,N_e]-[B_z,N_e]+\frac{2}{\beta_e}\nabla_\| \Delta_\perp A_\|=0\label{eq:gyro-2fields-Ne}\\
&&\partial_t (1-\frac{2\delta^2}{\beta_e}\Delta_\perp)A_\| -[\varphi,\frac{2\delta^2}{\beta_e}\Delta_\perp A_\|]+[B_z,\frac{2\delta^2}{\beta_e}\Delta_\perp A_\|] \nonumber\\
&& + \nabla_\| (\varphi-N_e-B_z)=0\label{eq:gyro-2fields-A}
\end{eqnarray}
with
\begin{eqnarray}
&&\left (\frac{2}{\beta_e}  +(1+2\tau)({\widetilde \Gamma}_0-{\widetilde \Gamma}_1) \right ) B_z=\nonumber \\
&&\left ( 1 -(\frac{{\widetilde \Gamma}_0-1}{\tau}) -{\widetilde \Gamma}_0 +{\widetilde \Gamma}_1  \right)\varphi  \label{gyro:Bzphi}\\
&&N_e=\left ( (\frac{{\widetilde \Gamma}_0-1}{\tau}) +\delta^2\Delta_\perp\right )\varphi\nonumber \\
&&-(1-{\widetilde \Gamma}_0+{\widetilde \Gamma}_1) B_z.\label{eq:gyro-Ne-phi}
\end{eqnarray}
Here,  ${\widetilde \Gamma}_n$ denotes the (non-local) operator 
$\Gamma_n(-\tau \Delta_\perp)$ associated to  the Fourier multiplyer  $\Gamma_n(\tau k_\perp^2)$, defined by  $\Gamma_n(x) = I_n(x) e^{-x}$ where $I_n$ is the modified Bessel function of first type of order n. 

In Eq. (\ref{eq:gyro-2fields-A}), the term $[B_z,\frac{2\delta^2}{\beta_e}\Delta_\perp A_\|]$ is sub-dominant in both scalings IV and V but, as mentioned above, it has been retained 
for it allows for a Hamiltonian formulation of the model in terms of a Lie-Poisson structure for the 2D limit, extended to 3D according to the procedure discussed in Ref. \cite{Tassi10}. We remark that the model and its Hamiltonian structure could also be derived from a drift-kinetic equation, by providing the relations (\ref{gyro:Bzphi})-(\ref{eq:gyro-Ne-phi}) and applying the procedure described in Ref. \cite{Tas15}.

We note also that the second term on the right-hand side of the relation (\ref{eq:gyro-Ne-phi}), which is proportional to $\delta^2$, corresponds to the above mentioned electron FLR correction, which is relevant when $\tau \beta_e \sim 1$.

\noindent
\textit{Remark{\tiny }:}
When neglecting the electon mass i.e. the $\delta^2$ contributions, expression (\ref{eq:gyro-Ne-phi}) for $N_e$ gives 
\begin{equation}
n_i = n_e = N_e + B_z=  \frac{{\widetilde \Gamma}_0 - 1}{\tau} \varphi 
+ ({\widetilde \Gamma}_0 -{\widetilde \Gamma}_1) B_z,
\end{equation}
consistent with Eq. (B1) of Ref. \cite{PS07}, originating from the 
low-frequency linear kinetic theory taken in the regime of adiabatic 
ions ($\zeta_i = \omega/(k_z v_{ti}) \gg 1$ and thus 
$R(\zeta_i) \ll 1$).

Substituting the expressions for $N_e$ and $B_z$  in Eqs. (\ref{eq:gyro-2fields-Ne})-(\ref{eq:gyro-2fields-A}), the resulting model only involves the electric and magnetic potentials $\varphi$ and $A_\|$. In the limit $\tau\ll~1$ 
where $N_e=\Delta_\perp\varphi$ and $B_z=-\frac{\beta_e/2}{1+\beta_e/2}\Delta_\perp\varphi$, and at large scales, where electron inertia can be neglected, one recovers Eqs. (3.2)-(3.3) and (3.10)-(3.12) of Ref. \cite{TSP16} (when taking the same assumptions mentioned at the beginning of the present Section). In this limit, it is possible to consider a finite value of $\beta_e$. 
If, on the other hand, electron inertia is kept into account, this system identifies (neglecting the subdominant term mentioned above) with the reduction to two fields (neglecting the coupling to $u_i$) of Eqs. (\ref{eq:FP1})-(\ref{eq:FP5}).  When $\beta_e$ is taken small enough so as to neglect  $B_z$ contributions,  Eqs. (\ref{eq:gyro-2fields-Ne})-(\ref{eq:gyro-2fields-A}) lead to the 2-field model of Refs. \cite{Schep94,Bor05}. 
\begin{eqnarray}
&&\partial_t\Delta_\perp \varphi+[\varphi,\Delta_\perp\varphi]+\frac{2}{\beta_e}\nabla_\|\Delta_\perp A_\|=0\label{eq:2fields-phi}\\
&&\partial_t(1-\frac{2\delta^2}{\beta_e}\Delta_\perp)A_\|-[\varphi,\frac{2\delta^2}{\beta_e}\Delta_\perp A_\|]\nonumber \\
&& +\nabla_\|(\varphi-\Delta_\perp\varphi)=0.\label{eq:2fields-A}
\end{eqnarray}

This model can also be derived from Eqs. (\ref{eq:uniform_lapphi})-(\ref{eq:uniform_Apar}) in the case $\tau=0$. It also 
corresponds the "low- $\beta$  case" of the two-fluid model of Ref. \cite{Biskamp97}
which restricts to 2D, when the electron pressure gradient in Ohm's law, usually 
referred as parallel electron compressibility 
(term  $\nabla_\| \Delta \varphi$  in Eq. (\ref{eq:2fields-A})) is not retained.

When $\tau\beta_e\sim 1$ one has (taking the limit $\tau\gg 1$), 
$B_z=\frac{\beta_e}{2} \varphi$ and $N_e=-\frac{\beta_e}{2} (1+\frac{2}{\beta_i}-\frac{2\delta^2}{\beta_e}\Delta_\perp)\varphi$, where $\beta_i$ denotes the ion beta parameter.
After neglecting subdominant corrections proportional to $\beta_e$, the system reduces to
\begin{eqnarray}
&&\partial_t (1+\frac{2}{\beta_i}-\frac{2\delta^2}{\beta_e}\Delta_\perp)\varphi-[\varphi,\frac{2\delta^2}{\beta_e}\Delta_\perp\varphi] \nonumber\\
&& -\frac{4}{\beta_e^2}\nabla_\|\Delta_\perp A_\|=0 \label{eq:KAW-tau-fini-phi}\\
&&\partial_t(1-\frac{2\delta^2}{\beta_e}\Delta_\perp)A_\|-[\varphi,\frac{2\delta^2}{\beta_e}\Delta_\perp A_\|]\nonumber \\
&&+\nabla_\|\varphi=0,\label{eq:KAW-tau-fini-A}
\end{eqnarray}
which identifies with the isothermal system  (5.9)-(5.10) of Ref. \cite{PST17} taken for large values of $\tau$ when  electron FLR corrections are neglected (see also Ref. \cite{Chen-Boldyrev17}).
This system also reproduces the "high-$\beta$ case" of Ref. \cite{Biskamp97} 
when restricted to 2D.

Similarly to many other reduced fluid and gyrofluid models (see Ref. \cite{Tas17} for a recent review), the system (\ref{eq:gyro-2fields-Ne})-(\ref{eq:gyro-2fields-A}), as above mentioned, possesses a noncanonical Hamiltonian structure. In order to show this point, we first observe that the system  (\ref{eq:gyro-2fields-Ne})-(\ref{eq:gyro-2fields-A}) can be formulated as an infinite-dimensional dynamical system with the fields $N_e$ and $A_e \equiv (1-2 \delta^2 \lapp /\beta_e)\apar$ as dynamical variables. Indeed, upon introducing the following positive definite operators 
\begin{eqnarray}
&&L_1 = \frac{2}{\beta_e}  +(1+2\tau)({\widetilde \Gamma}_0-{\widetilde \Gamma}_1)\\
&&L_2 = 1 +\frac{1-{\widetilde \Gamma}_0}{\tau} -{\widetilde \Gamma}_0 +{\widetilde \Gamma}_1 \\
&&L_3 = \frac{1-{\widetilde \Gamma}_0}{\tau} -\delta^2\Delta_\perp\\
&&L_4 = 1-{\widetilde \Gamma}_0+{\widetilde \Gamma}_1,
\end{eqnarray}
one can write $ B_z = M_1 \varphi$, with $M_1 =  L_1^{-1} L_2 $,
and $\varphi=-M_2^{-1}N_e$, where $M_2 = (L_3 + L_4L_1^{-1}L_2)$ is positive 
definite, as numerically seen on its Fourier transform. Also, $\apar=(1-2 \delta^2 \lapp /\beta_e)^{-1}A_e$. Thus, $B_z$, $\varphi$ and $\apar$ can be expressed in terms of the dynamical variables $N_e$ and $A_e$.

Proving that the system possesses a Hamiltonian structure amounts to show  that, given any observable $F$ of the system, i.e. a functional of $N_e$ and $A_e$, its evolution can be cast in the form \cite{Mor98}
\begin{equation}
\frac{\pa F}{\pa t}=\{F, \mathcal{E} \},
\end{equation}
where $\mathcal{E}$ is an observable corresponding to the Hamiltonian functional and $\{ \, , \, \}$ is a Poisson bracket. 

For the system (\ref{eq:gyro-2fields-Ne})-(\ref{eq:gyro-2fields-A}), the Hamiltonian is given by the conserved functional
\begin{eqnarray}
{\mathcal E} &=& \frac{1}{2} \int \Big ( \frac{2}{\beta_e} |\bnabla_\perp A_\| |^2 
+ \frac{4\delta^2}{\beta_e^2}|\Delta_\perp A_\| |^2 \nonumber \\
&-& N_e(\varphi -N_e-B_z) \Big ) d^3 {\boldsymbol x}, \label{energy}
\end{eqnarray}
whereas the Poisson bracket reads
\begin{align} 
&\{F,G\}=\int   \left(
( N_e ([F_{N_e} , G_{N_e}]+\delta^2 [F_{A_e} , G_{A_e}]) \right. \nno \\
&\left. +A_e([F_{N_e},G_{A_e}]+[F_{A_e} , G_{N_e}]) \right.  \nno \\
&\left. +F_{N_e}\pa_z G_{A_e} +F_{A_e}\pa_z G_{N_e} \right)  d^3 {\boldsymbol x}, \label{pb}
\end{align}
for two observables $F$ and $G$, and where subscripts on functionals denote functional derivatives. 

The Poisson bracket (\ref{pb}) corresponds, up to the normalization, to the Poisson bracket for the model of Ref. \cite{Schep94}, when the latter is reduced to a two-field model by setting the ion density fluctuations proportional to the vorticity fluctuations. As is common with noncanonical Hamiltonian systems \cite{Mor98}, the Poisson bracket (\ref{pb}) possesses Casimir invariants, corresponding to
\begin{equation}
C_{\pm}=\int  \, G_{\pm} d^3  {\boldsymbol x},
\end{equation}
where $G_\pm=A_e\pm\delta N_e$ are referred to as normal fields \cite{Wae09}. In terms of the normal fields,  the system (\ref{eq:gyro-2fields-Ne})-(\ref{eq:gyro-2fields-A}) rewrites in the form
\begin{equation}
\partial_t G_\pm +[\varphi_\pm,G_\pm]+\partial_z\left(\varphi_\pm  \mp \frac{1}{\delta}G_\pm  \right)=0,  \label{eq:Gpm}
\end{equation}
where $\varphi_\pm=\varphi-B_z\pm\frac{1}{\delta}A_\|$.

In the 2D limit with translational symmetry along $z$, the Poisson bracket takes 
the form of a direct product and the system possesses two infinite families of Casimir invariants, given by
\beq
C_{\pm}=\int  \mathcal{C}_{\pm} (G_\pm) d^2 \boldsymbol{x},
\eeq
with $\mathcal{C}_\pm$ arbitrary functions.  In particular, one has the quadratic invariants $\int G_\pm^2 d^2\boldsymbol{x}$, leading to the classical conservation of the magnetic potential in 2D MHD. In 2D, Eqs. (\ref{eq:Gpm}) take the form of advection equations for the Lagrangian invariants $G_{\pm}$ transported by incompressible velocity fields $\mathbf{v}_{\pm}=\hat{\boldsymbol{z}}\times \nabla \varphi_\pm$. Such Lagrangian invariants and velocity fields generalize those of the model of Ref. \cite{Caf98}.

We observe that the system admits also a further conserved quantity (which is not a Casimir invariant) corresponding to the generalized helicity
\begin{equation}
{\mathcal H} =\frac{1}{2} \int N_e \Big ( 1 - \frac{2\delta^2}{\beta_e}\Delta_\perp \Big) A_\|d^3 {\boldsymbol x}.
\end{equation}
This expression is similar (to dominant order) to the electron generalized helicity when 
making the assumptions $u_i=0$ and $\tau\ll 1$, where  $N_e$ then identifies to the vorticity.
The latter also rewrites 
\begin{equation}
{\mathcal H} =\frac{1}{8} \int (G_+^2-G_-^2) d^3 {\boldsymbol x}. \label{eq:hG+G-}
\end{equation}

At large scales, where $N_e = \Delta_\perp \varphi$ and $A_e = A_\|$, one has 
${\mathcal H} = -(1/2) \int \bnabla A_\|\bcdot \bnabla \varphi d^3{\boldsymbol x}
= (1/2) \int {\boldsymbol B}_\perp \bcdot {\boldsymbol u}_\perp  d^3{\boldsymbol x}$
which is the usual MHD cross-helicity.

\section{Phenomenology of critically-balanced KAW turbulence} \label{sec:KAW-turbulence}

In this section, we use the two-field gyrofluid model to phenomenologically characterize
the energy and/or helicity cascades which develop in strong KAWs turbulence. The aim is to predict 
the transverse magnetic energy spectrum together with the direct or inverse character of the cascades
in the different spectral ranges delimited by the plasma characteristic scales.

\subsection{Linear theory}


At the linear level, using a hat  to indicate  Fourier transform of fields and  Fourier symbols of operators,
one has the phase velocity $v_{ph}$ given by the dispersion relation 
\begin{equation}
v_{ph}^2\equiv\left(\frac{\omega}{k_z}\right)^2 = 
\frac{2}{\beta_e} \frac{k_\perp^2}{1 + \frac{2\delta^2 k_\perp^2}{\beta_e}} \frac{1 -{\widehat M}_1 + {\widehat M}_2} {{\widehat M}_2},
\end{equation}
where  $1 -{\widehat M}_1 + {\widehat M}_2$ is strictly positive for all $k_\perp$. 
\begin{figure}
	\centerline{
		\includegraphics[width=0.45\textwidth]{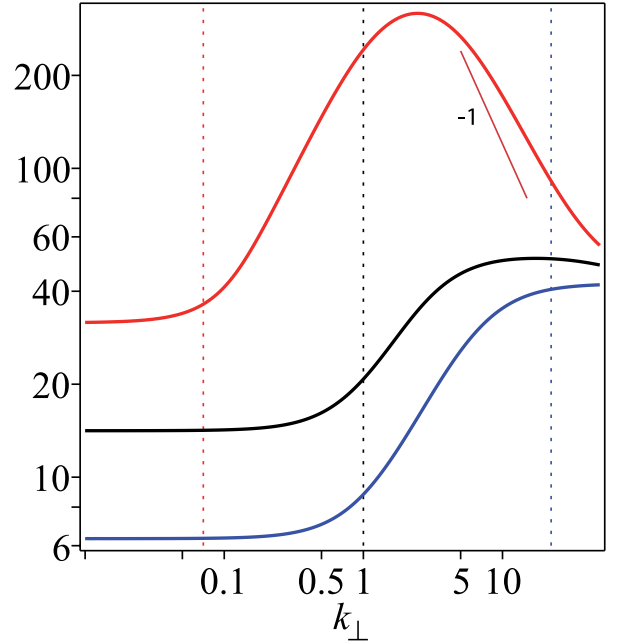}}
	\caption{Phase velocity of KAWs $v_{ph}$ versus $k_\perp$ for $\beta_e=0.002$, $\tau=100$ (red), $\beta_e=0.01$, $\tau=0.5$ (black) and  $\beta_e=0.05$, $\tau=0.001$ (blue). The vertical dotted lines refer to the inverse ion Larmor radius $\rho_i^{-1}$ for the three values of $\tau$, with the same color code as for $v_{ph}$. Transition between MHD and sub-ion 
    scales occurs at the smallest of the two scales $\rho_i$
    and $\rho_s$ (which corresponds to $k_\perp =1$). The orange straight line indicates the $k_\perp^{-1}$ asymptotic behavior in the large $\tau$ limit.}
	\label{fig1}
\end{figure}
The associated eigenmodes obey
\begin{equation}
{\widehat A}= \frac{\beta_e}{2} v_{ph} \frac{{\widehat M}_2} {k_\perp^2} {\widehat \varphi}.\label{eq:eigen}
\end{equation}
A graph of $v_{ph}(k_\perp)$ is displayed in Fig. \ref{fig1} for the cases $\beta_e=0.002$, $\tau=100$ (red), $\beta_e=0.01$, $\tau=0.5$ (black) and  $\beta_e=0.05$, $\tau=0.001$ (blue). An important difference that appears at large $\tau$, in addition to the shift  of the dispersive zone  towards smaller $k_\perp$ (due to the fact that $\rho_i$ is larger than $\rho_s$, here by a factor $\sqrt{200}$), is that at sub-$d_e$ scales, $v_{ph}$ does not stay constant but decreases as $k_\perp$ increases (asymptotically like $k_\perp^{-1}$ in the large $\tau$ limit), as in the full kinetic theory \cite{PST17}. In the absence of the $\delta^2$ term in $L_3$, $v_{ph}$ would be constant at small scales.

Interestingly, when assuming relation (\ref{eq:eigen}) in formula 
(\ref{energy}) for the energy ${\mathcal E}$, the sum of the first two terms of the energy ${\mathcal E}$ equals that of the last  three ones.

The magnetic compressibility $\chi = |{\widehat B}_z|^2 / |\widehat{{\boldsymbol B}}_\perp|^2$ associated with the Alfv\'en eigenmode is then given by
\begin{equation}
\chi =  \frac{2}{\beta_e} 
\left({1+ \frac{2\delta^2 k_\perp^2}{\beta_e}}\right)
\frac{{\widehat M_1}^2}{(1 -{\widehat M_1} +{\widehat M_2}){\widehat M_2}}.
\end{equation}	
\noindent
\textit {Small $\tau$ limit ($\tau \sim \beta_e^\frac{1}{2}$):} In this
regime, ${\widehat M}_1 \sim \beta_e k_\perp^2/2$ and is thus negligible (and so is $B_z$). On the other hand, ${\widehat M}_2 = 
(1-\Gamma_0)/\tau + O(\delta^2) \approx k_\perp^2 (1 - 3\tau k_\perp^2/4)$,
leading to the dispersion relation
  \begin{equation}
\left(\frac{\omega}{k_z}\right)^2 = 
\frac{2}{\beta_e} \frac{1}{1 + \frac{2\delta^2 k_\perp^2}{\beta_e}}
(1+ k_\perp^2 + \frac{3}{4}\tau k_\perp^2),
\end{equation}
consistent with the fluid formula given by Eq. (\ref{eq:dispKAW}).

\subsection{Absolute equilibria}

The invariants can be rewritten
\begin{eqnarray}
{\cal E} &=& \frac{1}{2} \int \Big [  (1- {\widehat M}_1 + {\widehat M}_2 ) {\widehat M}_2 |{\widehat \varphi}|^2 \nonumber \\
&&+ \frac{2k_\perp^2}{\beta_e}  \left (1 + \frac{2\delta^2k_\perp^2}{\beta_e} \right )|{\widehat A_\|}|^2\Big) 
\Big ]d^2 {\boldsymbol k}_\perp dk_z\\
{\cal H} &=& -\frac{1}{2} \int \Big [ {\widehat M}_2 \left ( 1 + \frac{2\delta^2 k_\perp^2 }{\beta_e^2} \right) ({\widehat \varphi}_R {\widehat A}_{\|R} +\widehat {\varphi_I}{\widehat A_{\| I}}) \Big ] \nonumber \\
&&d^2 {\boldsymbol k}_\perp dk_z
\end{eqnarray}
  with ${\widehat \varphi} ={\widehat \varphi}_R + i {\widehat \varphi}_I$ and $\widehat A_{\|}= \widehat A_{\| R} + i \widehat A_{\| I}$,  when separating real and imaginary parts.
  
Based on the existence of such quadratic invariants, a classical tool for predicting the direction of
  	turbulent cascades is provided by the behavior of the spectral density of the corresponding
  	invariants in the regime of absolute equilibrium. Albeit turbulence is intrinsically a non-equilibrium
  	regime and a turbulent spectrum strongly differs from an equilibrium spectrum, the increasing or
  	decreasing variation  of the latter in the considered spectral range can be viewed as reflecting
  	the direction of the turbulent transfer and thus the direct or inverse character of the cascade.
  	An early application of this approach to incompressible MHD is found in Ref. \cite{Pouquet75}. 
  	
 In order to apply equilibrium statistical mechanics to the system consisting in a finite number
  	of Fourier modes obtained by spectral truncation of the fields $A_\|$ and $\varphi$ governed
  	by Eqs. (\ref{eq:gyro-2fields-Ne}) and (\ref{eq:gyro-Ne-phi}), one first easily checks
  	that the solution satisfies the Liouville's theorem conditions in the form
  \begin{eqnarray}
  &&   \sum_{\boldsymbol k} \frac{\partial} {\partial {\widehat \varphi}_{R {\boldsymbol k}}}
  \left (\frac{\partial  {\widehat \varphi}_{R {\boldsymbol k}}}{\partial t} \right )
  + \frac{\partial} {\partial {\widehat \varphi}_{I {\boldsymbol k}}}
  \left (\frac{\partial  {\widehat \varphi}_{I {\boldsymbol k}}}{\partial t} \right )=0\\
  && \sum_{\boldsymbol k} \frac{\partial} {\partial {\widehat A}_{\|R {\boldsymbol k}}}
  \left (\frac{\partial  {\widehat A}_{\|R {\boldsymbol k}}}{\partial t} \right )
  + \frac{\partial} {\partial {\widehat A}_{\|I {\boldsymbol k}}}
  \left (\frac{\partial  {\widehat A}_{\| {\boldsymbol k}}}{\partial t} \right )=0. 
  \end{eqnarray}
    
 The density in phase space of the canonical equilibrium ensembles for the system (\ref{eq:gyro-2fields-Ne})-(\ref{eq:gyro-2fields-A}), truncated in Fourier space,  is given by
 $\rho=Z^{-1}\exp(-\lambda \mathcal{E}-\mu \mathcal{H})=Z^{-1}\exp(-M_{ij}x^ix^j/2)$, where $Z$ is the partition function. The matrix $M$ is defined as
  $$
  \quad
  {\mathbf M} =\begin{bmatrix}
  	f & 0 & h & 0 \\
  	0 & f & 0 & h \\
  	h & 0 & g & 0 \\
  	0 & h & 0 & g
  \end{bmatrix}
  $$
  where $ f = \lambda (1- {\widehat M}_1 + {\widehat M}_2 ) {\widehat M}_2$, $ g = \lambda \frac{2k_\perp^2}{\beta_e}  \left (1 + \frac{2\delta^2k_\perp^2}{\beta_e} \right )$ and $h = \frac{\mu}{2}
   {\widehat M}_2 \left ( 1 + \frac{2\delta^2 k_\perp^2 }{\beta_e^2} \right)$. Here,
   $\lambda$ and $\mu$ denote numerical constants prescribed by the values of the total energy and helicity.
The symbols $x^i_{i=1,4}$ refer to  $ {\widehat \varphi}_R$, $ {\widehat \varphi}_I$,  $\widehat A_{\| R}$ and $\widehat A_{\| I}$.
The inverse matrix easily writes   
   
  $$
  \quad
  {\mathbf M}^{-1} = \frac{1}{\Delta}\begin{bmatrix}
  g & 0 & -h & 0 \\
  0 & g & 0 & -h \\
  -h & 0 & f & 0 \\
  0 & -h & 0 & f
  \end{bmatrix},
  $$ 
 with $\Delta = fg-h^2$. Without dissipation, the statistical equilibrium has an energy spectral density 
 \begin{equation}
 E_k \sim \frac{1}{\lambda}2 \pi k (f E^\varphi_k + g E_k^{A_\|})
 \end{equation}
 and a helicity spectral density 
 \begin{equation}
 H_k \sim \frac{1}{\mu} 4 \pi k_\perp h E_k^{\varphi A_\|},
 \end{equation}
 where $E_k^{\varphi}= g/\Delta$, $E_k^{A_\|} =f /\Delta$ and $E_k^{\varphi A_ \|} = -h/\Delta$.
 
 The cascade directions are forward or backward, depending on whether the absolute equilibrium spectra are respectively growing or decreasing in the wavenumber ranges of interest.
 The energy spectrum rewrites
 \begin{equation}
 E_k\sim \frac{4\pi}{\lambda}\frac{k_\perp}{1-\frac{\mu^2}{4\lambda^2 }\frac{1}{v_{ph}^2}}.
 \end{equation}
 Positivity condition prescribes constraints on the wavenumber domain where this formula applies. The condition  $\mu/\lambda\lesssim 2 \min (v_{ph})$  (where $\min(v_{ph})=\sqrt{8/\beta_e}$  for small $\tau$ but is smaller for larger values of $\tau$), ensures that the energy spectrum is defined for all wavenumbers. 
 For larger values of $\mu/\lambda$, there is a lower bound in $k_\perp$ and possibly also an upper bound, for which $E_k>0$. As $v_{ph}$ is bounded from above, it might happen that the energy is never positive. 
 A more detailed study would require to explicitly relate the constants $\mu$ and $\lambda$ to the total energy and helicity.
 Nevertheless, in all the cases where it is defined, the energy is found to be a growing function of $k_\perp$ (except possibly near the lower $k_\perp$ bound where it has a singular behavior), whatever the values of $\beta_e$ and $\tau$, indicating a forward cascade.
 The generalized helicity spectrum, on the other hand, rewrites
 \begin{equation}
  H_k\sim -\frac{4\pi}{\mu}\frac{k_\perp}{\frac{4\lambda^2}{\mu^2}v_{ph}^2-1},
 \end{equation}
 which is negative. We thus have the relation $H_k=-\mu/(4\lambda) E_k/v_{ph}^2$. 
 Note however that there is no definite sign for this spectrum.
In the same wavenumber ranges where the energy is positive, its absolute value is a growing quantity both at MHD and sub-$d_e$ scales. However, in the intermediate 
 (sub-$\rho_s$ or sub-$\rho_i$) range, where $\omega/k_z \sim k_\perp$, it is a decreasing function of $k_\perp$, indicating an inverse cascade. Note that when the $-7/3$ power law of the turbulent transverse magnetic energy spectrum is not well developed (see next Section), the range of generalized helicity inverse cascade is also very limited. Similar results showing an inverse (or direct) helicity cascade in the Hall (respectively sub-electronic) range  are obtained in Ref. \cite{Mil17} based on absolute equilibrium arguments in extended MHD (XMHD).

 \subsection{Turbulent spectra}
 
 \subsubsection{Energy cascade}
 We here discuss the turbulent state  in the presence of a small amount of dissipation at small scales (leading to a finite flux of energy), focusing on the case of a critically balanced KAW cascade (with equal amount of positively and negatively propagating waves).
 Following the discussion of Section 7 in Ref. \cite{PST17}, the magnetic spectrum is easily obtained by imposing a constant energy flux, estimated by ratio of the spectral energy density 
 at a given scale by the nonlinear transfer time at this scale. In the strong wave (critically-balanced) turbulence regime, this energy transfer time reduces to the nonlinear timescale. To estimate these quantities, it is first  necessary to relate the Fourier components of the electric and magnetic potentials. This is achieved assuming the linear relationship provided by Eq. (\ref{eq:eigen}), characteristic of Alfv\'en modes.
 After inserting this relation into the energy ${\mathcal{E}}$ one finds that the total 3D spectral energy density writes
 \begin{equation}
 {\mathcal E}^{3D}_{{\boldsymbol k}}=\frac{2}{\beta_e}k_\perp^2 \left( 1+\frac{2\delta^2k_\perp^2}{\beta_e}\right)|{\widehat A}_{\boldsymbol k}|^2
 \end{equation}
 Due to the quasi-2D character of the dynamics, it is convenient to deal with the 2D energy spectrum 
 \begin{equation}
 {\mathcal{E}^{2D}_k}=\frac{2}{\beta_e}k_\perp^2 \left( 1+\frac{2\delta^2k_\perp^2}{\beta_e}\right)|{\widehat A}_{k_\perp}|^2
 \end{equation}
 where we used the notation
 \begin{equation}
 |{\widehat A}_{k_\perp}|^2 = \int |{\widehat A}_{\boldsymbol k}|^2 dk_z,
 \end{equation}
 and assume statistical isotropy in the transverse plane. 
 Similar definitions are used for the other relevant fields, 
 namely the electrostatic potential $\varphi$ and the transverse
 magnetic field ${\boldsymbol B}_\perp$.
 
The nonlinear timescale is estimated from Eq. (\ref{eq:gyro-2fields-A}) which, after discarding the $B_z$ terms (smaller by a factor $\beta_e$) and the $\partial_z$ terms, can be rewritten
 \begin{equation}
 \partial_t A_e+[\varphi,A_e] -[A_\|, M_2\varphi]=0.
 \end{equation} 
 Assuming locality of the nonlinear interactions in Fourier space, the typical frequencies at wavenumber $k_\perp$ associated with the two nonlinear terms of the above equation take the form
 $\tau_{NL1}^{-1}({k_\perp})\sim k_\perp^2 |\widehat{\varphi}_{k_\perp}|$ and $\tau_{NL2}^{-1}({k_\perp}) \sim k_\perp^2 {\widehat M}_2|\widehat{\varphi}_{k_\perp}|/(1+2\delta^2k_\perp^2/\beta_e)$ respectively. The global nonlinear frequency of the system can be estimated by a linear combination of these two frequencies. Taking equal weights leads to the estimate
 \begin{equation}
 \tau_{NL}^{-1}({k_\perp})\sim\frac{2}{\beta_e}k_\perp^4\left (1+\frac{\widehat{M}_2}{1+\frac{2\delta^2k_\perp^2}{\beta_e}} \right )\frac{1}{\widehat{M}_2v_{ph}}|\widehat{A}_{k_\perp}|.
 \end{equation}
 In two-dimensions, when assuming isotropy, the transverse magnetic energy spectral density $|\widehat{B_\perp}({k_\perp})|^2 \sim
 k_{\perp}^2 |{\widehat A}_{k_\perp}|^2$ is  related to the transverse magnetic energy spectrum by
 $E_{B_\perp}(k_\perp)\sim k_\perp^{-1} |{\widehat B}_\perp(k_\perp)|^2$, the energy flux $\varepsilon$ writes
 \begin{equation}
 \varepsilon\sim\frac{4}{\beta_e^2}\left (1+\frac{2\delta^2k_\perp^2}{\beta_e}+\widehat{M}_2 \right )\frac{1}{\widehat{M}_2 v_{ph}} k_\perp^3|\widehat{B}_\perp({k_\perp})|^3, 
 \end{equation}
 and thus, assuming a constant energy flux, one gets
 \begin{equation}
 E_{B_\perp}(k_\perp)\sim \varepsilon^{2/3}\beta_e^{4/3}k_\perp^{-3}\left ( \frac{v_{ph}\widehat{M}_2}{1+\frac{2\delta^2k_\perp^2}{\beta_e}+\widehat{M}_2} \right )^{2/3}.\label{eq:Espectrum}
 \end{equation}
 
 All the regimes of KAW energy cascade can be recovered from Eq. (\ref{eq:Espectrum}).
 
$\bullet$ {\it MHD range}

At scales large compared to $\rho_s$ and $\rho_i$, one has $v_{ph}\sim (2/\beta_e)^{1/2}$, $\widehat{M}_2=k_\perp^2$ and $k\ll 1$. One thus immediately finds  $E_B(k)\sim \varepsilon^{2/3}k_\perp^{-5/3}$.
 
$\bullet$ {\it Sub-$\rho_i$ range}

 When $\sqrt{\beta_e/2}/\delta\ge k_\perp  \gtrsim (2\tau)^{-1/2}$ and $\tau\ge 1$ (i.e. for scales smaller than the ion gyroradius (assumed larger than $\rho_s$), for which $\Gamma_0 \approx 0$ and $\Gamma_1 \approx 0$, and large enough for electron inertia to be negligible), one has $\widehat{M}_2\sim 1/\tau+\beta_e(1+\tau)/(2\tau)\sim {\rm constant}$ and $v_{ph}\sim k_\perp$, so that $E_B(k)\sim \varepsilon^{2/3}k_\perp^{-7/3}$. 

$\bullet$ {\it Sub-$\rho_s$ range}

 When, on the other hand, $\tau\le 1$, for 
scales intermediate between $\rho_s$ and $d_e$, characterized by $k_\perp\gg 1$ and  $2\delta^2 k_\perp^2/\beta_e  \ll 1$, one finds $\widehat{M}_2\sim k_\perp^2$ and  $v_{ph}\sim (2/\beta_e)^{1/2} k_\perp$, so that again 
 $E_{B_\perp}(k_\perp)\sim \varepsilon^{2/3}k_\perp^{-7/3}$. It is however to be noted that in this case, the smallest nonlinear time scale is not the stretching time $\tau_{NL1}$ but rather $\tau_{NL2}$, associated with the electron pressure term
 in Ohm's law or equivalently to the Hall term, as previously mentioned.
 
$\bullet$ {\it Sub-$d_e$ range} 

When $\beta_e$ is small enough, it is possible to observe a third power law at scales smaller that the electron inertial length (but still larger than the electron Larmor radius). 

${\bf -}$  When $\tau \ll 1$, the $-7/3$ power-law zone is almost inexistent. It is replaced by  a smooth transition between the $-5/3$ power-law and a steeper zone where $v_{ph}\sim {\rm cst}$, $\widehat{M}_2\sim k_\perp^2$ and thus where  $E_B(k)\sim \varepsilon^{2/3}k_\perp^{-3}$. 

${\bf -}$ If $\tau$, is taken larger than unity, $v_{ph}\sim k_\perp^{-1}$ and $\widehat{M}_2\sim k_\perp^2$, leading to $E_{B_\perp}(k_\perp)\sim \varepsilon^{2/3}k_\perp^{-11/3}$. 
 
${\bf -}$ Note that for a small range of parameters where $\beta_e \ll 1$ and $\tau=O(1)$, a regime where one can have $v_{ph}\sim {\rm constant}$ and  $\widehat{M}_2\sim {\rm constant}$, one recovers a spectrum
of the form $E_{B_\perp}(k_\perp)\sim \varepsilon^{2/3}k_\perp^{-13/3}$, as mentioned in Ref. \cite{PST17}.
 
 \subsubsection{Generalized helicity cascade}
We here derive the expected transverse  magnetic energy spectrum associated with a generalized helicity cascade. 
Proceeding as in the case of the energy cascade, we first write the 3D spectral density (taken positive)
\begin{equation}
{\cal H}^{3D}_k=\frac{1}{\beta_e v_{ph}}(1+\frac{2\delta^2k^2}{\beta_e})k_\perp^2|\widehat{A}_{\boldsymbol k}|^2.
\end{equation}
Keeping the same estimate for the transfer time, and assuming a constant generalized helicity flux rate $\eta$, we obtain the magnetic spectrum in the helicity cascade
\begin{equation}
E_{B_\perp}(k_\perp)\sim \eta^{2/3}\beta_e^{4/3}k_\perp^{-3}\left ( \frac{v_{ph}^2\widehat{M}_2}{1+\frac{2\delta^2k_\perp^2}{\beta_e}+\widehat{M}_2} \right )^{2/3}.\label{eq:Hspectrum}
\end{equation}
Going through the same estimates in the various wavenumber domains as for the energy cascade, we now see that the magnetic spectrum in the helicity cascade obeys a $-5/3$ power law from the MHD range to the electron scale. At scales smaller that $d_e$, we differently finds that for $\tau\ll 1$ the spectrum is proportional to $k_\perp^{-3}$, while it is otherwise  proportional to $k_\perp^{-13/3}$.

It is of interest to remark that this latter scaling is somewhat similar to the $M_{H+}$ spectrum of \cite{Abdelhamid16} associated to the magnetic spectrum of the magneto-sonic cyclotron branch in the so-called H-generalized helicity cascade computed on exact solutions of an extended MHD model (with the caveat that in \cite{Abdelhamid16} a singularity appears at the $d_e$ scale).

\begin{figure}
	\centerline{
		\includegraphics[width=0.45\textwidth]{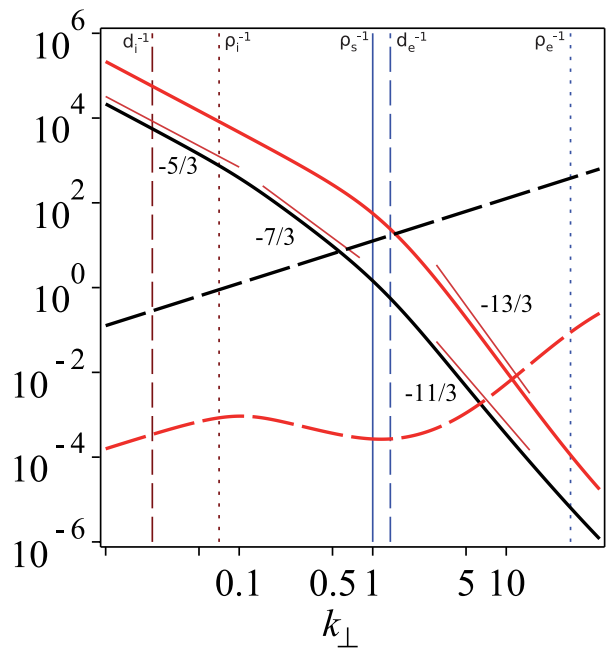}}
		\caption{Turbulent magnetic spectra (solid lines) in energy (black) and generalized helicity (red) cascades, together with absolute equilibrium energy (black long dashed lines) and generalized helicity (red long dashed lines) spectra for $\beta_e=0.002$, $\tau=100$. Straight orange lines refer to the slopes of the various power-law inertial ranges: $-5/3$ in the MHD range, $-7/3$ in the sub-ion Larmor radius range and $-11/3$ (for the energy cascade) or $-13/3$ (for the helicity cascade) in the sub-$d_e$ range. The blue solid vertical line refers to $\rho_s^{-1}$, the brown and blue long-dashed (respectively dotted) vertical lines  to the inverse ion and electron inertial lengths (respectively Larmor radii) $d_i^{-1}$ and $d_e^{-1}$ (respectively $\rho_i^{-1}$ and $\rho_e^{-1}$).}
	\label{fig2}
\end{figure}
\begin{figure}
	\centerline{
		\includegraphics[width=0.45\textwidth]{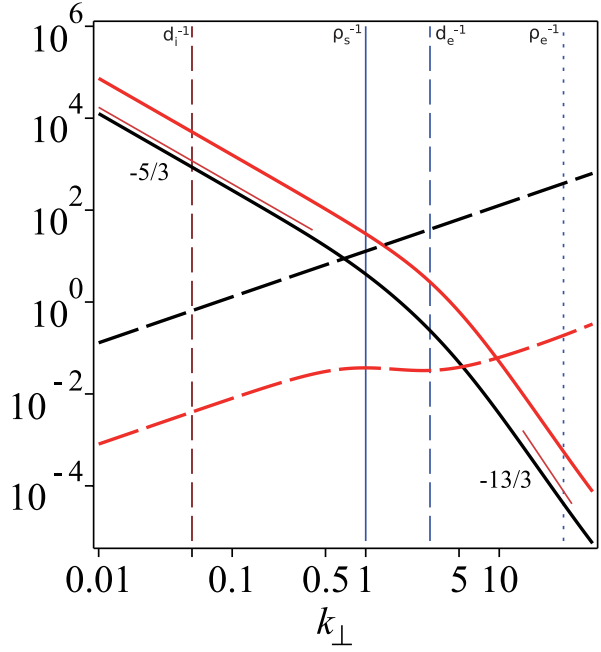}}
	\caption{Same as for Fig. \ref{fig2} for $\beta_e=0.01$, $\tau=0.5$. No sub-ion power-law range is visible. Both for the energy and helicity cascades, the magnetic spectrum displays a $-13/3$ sub-$d_e$ power-law range.}
	\label{fig3}
\end{figure}
\begin{figure}
	\centerline{
		\includegraphics[width=0.45\textwidth]{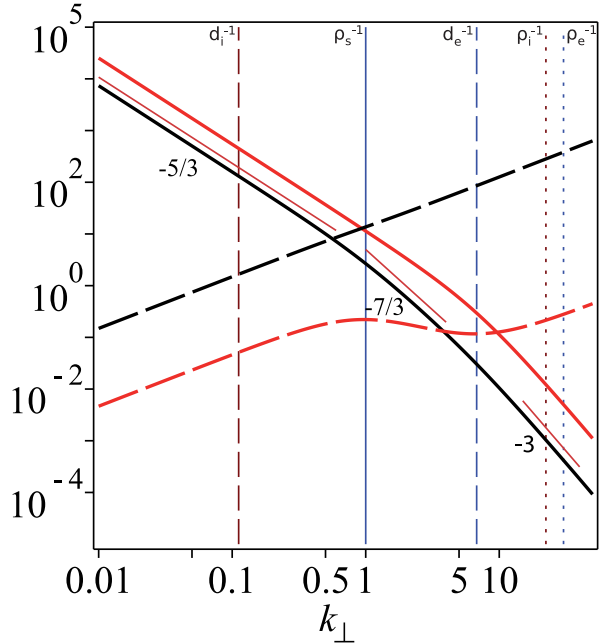}}
	\caption{Same as for Fig. \ref{fig2} for $\beta_e=0.05$, $\tau=0.001$. A $-7/3$ power-law turbulent magnetic spectrum in the energy cascade is visible for $k_\perp >1$, while, both for the energy and helicity cascades, the magnetic spectrum displays a $-3$ sub-$d_e$ power-law range.}
	\label{fig4}
\end{figure}

Examples of transverse magnetic energy spectra are displayed for the parameters $\beta_e=0.002$, $\tau=100$ (Fig.  \ref{fig2}), $\beta_e=0.01$, $\tau=0.5$ (Fig.  \ref{fig3}) and $\beta_e=0.05$, $\tau=0.001$ (Fig. \ref{fig4}), both for the absolute equilibria (long dashed lines) of the energy (black) and the generalized helicity (red) and for the turbulent magnetic spectra (solid lines) associated to the energy cascade (black) and the helicity cascade (red). The helicity inverse cascade  associated with the decreasing absolute equilibrium spectrum in sub-ion range, is conspicuous in the  case of large $\tau$, but less pronounced for $\tau$ of order unity.

\section{Discussion and conclusion}\label{sec:conclusion}

In this paper, two new reduced models have been derived for 
low-$\beta_e$ plasmas. 
One of them, given by Eqs. (\ref{eq:uniform_lapphi})-(\ref{varphistar}), concerns the small
$\tau$-regime and extends the four-field model of Ref. \cite{Hsu86} by
retaining electron inertia. Both a fluid derivation and a reduction of the gyrofluid model of Ref. \cite{Bri92} are presented. Interestingly, agreement between the two
formulations requires  closure  assumptions consistent with the underlying 
scaling, such as adiabatic ions. The other 
model, given by Eqs. (\ref{eq:KAW-tau-fini-phi})-(\ref{eq:KAW-tau-fini-A}), is a two-field gyrofluid model, valid for any $\tau$, which 
retains both electron inertia and $B_z$ fluctuations, in addition to ion FLR 
contributions. It is used to present a comprehensive phenomenological description of the Alfv\'en wave magnetic energy spectrum from the MHD
scales to scales smaller than $d_e$ (while larger than $\rho_e$). Assuming the existence of energy or helicity cascades, this leads  to the prediction of the magnetic energy spectrum when neglecting possible intermittency effects originating from the presence of coherent structures. 
The existence of these cascades needs to be confirmed by numerical simulations of the gyrofluid equations supplemented by dissipation and energy and/or helicity injection. In particular, the inverse helicity cascade is expected to occur only when the system is driven at a scale close to $d_e$, in a way that mostly injects helicity rather than energy. In fact, Eq. (\ref{eq:hG+G-}) shows that a non-zero helicity corresponds to an imbalanced regime where either $G_+$ or $G_-$ dominates.  It is interesting to note that the evidence of an inverse helicity cascade in numerical simulations of imbalanced EMHD turbulence was reported  in Refs. \cite{Cho16, KimCho15}. Analytic considerations on the role of helicity in weak REMHD turbulence can also be found in Ref. \cite{Galtier15}. An imbalanced energy injection could possibly originate from magnetic reconnection that takes place at the electronic scales. This scenario was recently considered in Ref. \cite{Franci17} on the basis of 2D hybrid PIC and Vlasov simulations where the development of a sub-ion magnetic energy spectrum occurs in relation with the reconnection instability, before the direct energy cascade reaches this scale.

In the framework of the two-field gyrofluid model, the transition scale between the $k_\perp^{-5/3}$ and the  $k_\perp^{-7/3}$ ranges occurs at 
the largest of the two scales $\rho_i$ and $\rho_s$. When $\tau$
is small, this will also be the case with  Eqs. (\ref{eq:uniform_lapphi})-(\ref{varphistar}) that retain the coupling to $n= -(2/\beta_e) B_z$ and $u_i$, as shown by using the same arguments as in Appendix
E.4 of Ref. \cite{Schekochihin09}. Differently for $\tau \sim 1$ and small $\beta_e$, 
a spectral transition is observed to take place at scale $d_i$, 
both  in the solar wind \cite{Chen14} and in hybrid-PIC simulations \cite{Franci16}.
The question arises whether a similar transition could also be observed in numerical
simulations of reduced models, induced by the presence of current sheets and the occurence of
reconnection processes, or if more physics has to be taken into account.

Note that while the magnetic energy spectrum  displays a $k_\perp^{-7/3}$ range both below the ion Larmor radius and below $\rho_s$ when $\beta_e$ is small, the perpendicular 
electric field spectrum scales like $k_\perp^{-1/3}$ in the former regime and like $k_\perp^{-13/3}$ in the latter one. 

The two-field gyrofluid model derived in this paper could be extended to account for electron Landau damping, a crucial ingredient at small $\beta_e$, with either a Landau fluid formulation, as suggested in Ref. \cite{PST17}, or with the coupling with a drift-kinetic equation. In the latter case, it could provide an interesting generalization of the model presented in \cite{ZS11}, by taking into account the parallel magnetic field fluctuations and thus permitting larger values of $\beta_e$.

At sub-$d_e$ scales, a new regime is uncovered in the case of cold ions (small $\tau$),
where the magnetic energy density scales like $k_\perp^{-3}$. Compressibility here plays a 
central role, which explains the difference with the cases $\tau \sim 1$ where the
spectrum scales like $k_\perp^{-13/3}$  or $\tau \gg 1$ (a quasi-incompressible limit)
where it scales like $k_\perp^{-11/3}$. Scales smaller than $\rho_e$ are
not considered in this paper, as they require a full description of the electron FLR effects.
In this regime, the spectrum is observed to be even steeper \cite{Huang14}, possibly associated
with a phase-space entropy cascade \cite{Schekochihin09}.

We have here considered the regime of strong wave turbulence
where critical balance holds. Due to this property, the estimates of the nonlinear times and the relation 
between the fields turn out to be identical to those
of the purely non-linear regime that occurs
for example in two dimensions.

{\bf Acknowledgments:} We are thankful to W. Dorland for useful discussions.

\appendix 

\section{Dispersion relation} \label{app:dispersion}
The system (\ref{eq:uniform_lapphi})-(\ref{eq:uniform_Apar}), when linearized about a uniform state, leads to
\begin{eqnarray}
&& \frac{\omega}{k_z}(1+\frac{5\tau}{4}k_\perp^2)\widehat{\varphi^*}-\frac{2}{\beta_e}(1+\frac{\tau}{2}k_\perp^2)\widehat{A_\|}=0
\label{eq:linear1}\\
&&-\frac{\omega}{k_z}((1+\delta^2+\tau k_\perp^2)\widehat{u_i}
-\frac{2\delta^2}{\beta_e}k_\perp^2\widehat{A_\|})\nonumber \\ 
&& +(1+\tau) \widehat{n}+\tau k_\perp^2\widehat{\varphi^*}=0 \\
&& -\frac{\omega}{k_z} \widehat{n} + \widehat{u_i}-\frac{2}{\beta_e}k_\perp^2\widehat{A_\|}=0\\
&&-\frac{\omega}{k_z}((1+\frac{2\delta^2}{\beta_e}k_\perp^2)
\widehat{A_\|}-\delta^2\widehat{u_i} )+(1+\frac{\tau}{2}k_\perp^2)\widehat{\varphi^*}\nonumber \\
&& -(1+\tau) \widehat{n}=0, 
\end{eqnarray}
where $\omega$, $k_z$, $k_\perp$ are respectively the frequency, parallel and perpendicular wavenumbers of harmonic perturbations
whose Fourier complex coefficients are denoted with a $\widehat{.}$ symbol.
This system supports two kind of waves, kinetic Alfv\'en waves (KAWs) and slow-magnetosonic waves (SWs). Ion parallel velocity plays a minor role in the dispersion relation of KAWs that can thus be approximated by
\begin{equation}
(\frac{\omega}{k_z})^2=\frac{2}{\beta_e}\frac{1}{1+\frac{2\delta^2k_\perp^2}{\beta_e}}\left((1+\tau)k_\perp^2+\frac{(1+\tau k_\perp^2/2)^2}{1+5\tau k_\perp^2/4}\right).\label{eq:dispKAW}
\end{equation}
It turns out that this approximation is excellent for a wide range of values of $\tau$ and $\beta_e$ in the whole spectral domain.  Another simplification consisting in taking the cold ion limit and dropping some subdominant contributions proportional to $\delta^2$, allows one to obtain the slow branch. The dispersion relation then reduces to
\begin{equation}
\Big((\frac{\omega}{k_z})^2-1\Big )\Big((\frac{\omega}{k_z})^2-\frac{2}{\beta_e}\Big)+\frac{2k_\perp^2}{\beta_e}(\frac{\omega}{k_z})^2\Big(\delta^2(\frac{\omega}{k_z})^2 -1\Big)=0.\label{eq:disptauO}
\end{equation}
It is easy to verify that the KAW dispersion relation given in Eq. (\ref{eq:dispKAW}) taken for $\tau=0$, can be recovered from Eq. (\ref{eq:disptauO}) when $\frac{\omega}{k_z}\gg 1$. The slow magnetosonic branch is such that $\frac{\omega}{k_z}\sim 1$ at large scale, with a small dispersive component at small scale (a good approximation to the solution is given by $\omega/k_z=(1+k_\perp^2)^{-1/2}$).
  From these results, one can estimate, for both kinds of waves and within scaling II, the values of $\zeta_r=\omega/(k_z v_{th r})$ both for ions (for which $v_{th i}\sim \tau^{1/2}\sim \delta^{1/2}$) and for electrons (for which $v_{th e}\sim \delta^{-1}$). On has, for KAWs, $\zeta_i\sim \delta^{-3/2}\gg 1$ and $\zeta_e \sim 1$, while for SWs, $\zeta_{i}\sim \delta^{-1/2}\gg 1$ and $\zeta_e\sim \delta \ll 1$. It is thus a reasonable approximation to assume adiabatic ions and isothermal electrons. The good agreement between kinetic theory and an isothermal equation of state for the electrons, even when $\zeta_e\sim 1$, is shown in Ref. \cite{PST17}.

\section{Parent gyrofluid model}  \label{app:gyrofluid}

We adopt the same definitions of Ref. \cite{TSP16} and consider the following gyrofluid equations for the evolutions of the gyrocenter moments
$N_{e,i}$, $U_{e,i}$, $P_{\parallel e,i}$, $P_{\perp e,i}$, $Q_{\parallel e,i}$, $Q_{\perp e,i}$, $R_{\parallel \perp e,i}$ and $R_{\perp \perp e,i}$ corresponding to the the normalized fluctuations of gyrocenter density, parallel velocity, parallel and perpendicular pressure, parallel and perpendicular heat flux, and of the parallel/parallel and parallel/perpendicular components of the energy weighted pressure tensor respectively, with the subscript $e$ and $i$ referring to electrons and ions
\begin{align}
&\frac{\pa N_e}{\pa t}+[\ede \varphi , N_e]+\dd[ \ds \ede \varphi,  \Ppee - N_e]-[ \ede \apar , U_e] \nno \\
&-[\ede \bpar , \Ppee] -\dd [\ds \ede \bpar , \Ppee - N_e]+\frac{\pa U_e}{\pa z}=0,   \label{neg}\\
&\frac{\pa }{\pa t} \left(\dd U_e - \ede \apar \right) + \dd [\ede \varphi ,  U_e] -[\ede \apar , \Ppe] \nno \\
&-\dd [\ds  \ede \apar , \Ppee - N_e] - \dd [ \ede \bpar , U_e]- \dd [\bpar , \Qpee] \nno \\
&-{\overline \Gamma}_0 (\dd \ds^{\varphi} , \dd \ds^{A})[\varphi , \apar] \nonumber\\
&+({\overline \Gamma}_0 (\dd \ds^B , \dd \ds^A)+\dd \ds {\overline \Gamma}_1 (\dd \ds^B , \dd \ds^A))[\bpar , \apar] \nonumber \\
&+\frac{\pa}{\pa z}\left(\Ppe-\ede \varphi + \ede \bpar\right)=0,   \label{ueg}\\ 
&\frac{\pa \Ppe}{\pa t}+[\ede \varphi , \Ppe]+\dd [\ds \ede \varphi , \Ppee - N_e] -2 [  \apar , \Qpe] \nno \\
&- 3 [ \ede \apar , U_e] - [ \ede \bpar , \Ppe + \Ppee -N_e] \nno \\
& - \dd [ \ds  \ede  \bpar , \Ppee - N_e] - [\bpar , \Rxe] \nonumber \\
& + \frac{\pa }{\pa z}\left( \Qpe + 3 U_e\right)=0, \label{peg}\\
&\frac{\pa \Ppee}{\pa t}+[(1+ \dd \ds)\ede \varphi  , \Ppee] \nno \\
& + \dd [\ds (2 + \dd \ds) \ede \varphi , \Ppee - N_e] - [\ede \apar  , U_e] \nonumber \\
&-[\apar , \Qpee] -[(2+\dd \ds)\ede \bpar , 2 \Ppee - N_e] \nno \\
&- \dd [\ds (3 + \dd \ds )\ede  \bpar , \Ppee - N_e] \nonumber \\
&-2[\bpar , \Rpe]+\frac{\pa}{\pa z}\left(U_e + \Qpee\right)=0, \label{ppeg}\\
&\frac{\pa N_i}{\pa t}+[\edi \varphi , N_i]+\tau[ \ds \edi \varphi,  \Ppei - N_i]-[ \edi \apar , U_i] \nno \\
&+\tau [\edi \bpar , \Ppei] +\tau^2 [\ds \edi \bpar , \Ppei - N_i]+\frac{\pa U_i}{\pa z}=0,  \label{nig} \\
&\frac{\pa }{\pa t} \left( U_i + \edi \apar \right) +  [\edi \varphi ,  U_i] -\tau [\edi \apar , \Ppi]  \nno \\
&-\tau^2 [\ds  \edi \apar , \Ppei - N_i] + \tau [ \edi \bpar , U_i]+ \tau [\bpar , \Qpei]  \nno \\
&+ {\overline \Gamma}_0 (\tau \ds^{\varphi} , \tau \ds^{A})[\varphi , \apar] \nonumber \\
&+\tau ({\overline \Gamma}_0 (\tau \ds^B , \tau \ds^A)+\tau \ds 
{\overline \Gamma}_1 (\tau \ds^B , \tau \ds^A))[\bpar , \apar] \nno \\
&+\frac{\pa}{\pa z}\left(\tau \Ppi+\edi \varphi + \tau\edi \bpar\right)=0,  \label{uig}\\ 
&\frac{\pa \Ppi}{\pa t}+[\edi \varphi , \Ppi]+\tau [\ds \edi \varphi , \Ppei - N_i] -2 [  \apar , \Qpi] \nno \\
&- 3 [ \edi \apar , U_i] + \tau[ \edi \bpar , \Ppi + \Ppei -N_i] \nno \\
&   + \tau^2 [ \ds  \edi  \bpar , \Ppei - N_i] + \tau [\bpar , \Rxi] \nno \\
&+ \frac{\pa }{\pa z}\left( \Qpi + 3 U_i\right)=0, \label{pig}\\
&\frac{\pa \Ppei}{\pa t}+[(1+ \tau \ds)\edi \varphi  , \Ppei] \nno\\
& + \tau [\ds (2 + \tau \ds) \edi \varphi , \Ppei - N_i] \nno \\
&- [\edi \apar  , U_i]  -[\apar , \Qpei] \nno\\
& +\tau[(2+\tau \ds)\edi \bpar , 2 \Ppei - N_i] \nno \\
&+ \tau^2 [\ds (3 + \tau \ds )\edi  \bpar , \Ppei - N_i]  \nonumber \\
&+2\tau[\bpar , \Rpi]+\frac{\pa}{\pa z}\left(U_i + \Qpei\right)=0,  \label{ppig}
\end{align}
together with Poisson's equations and parallel and perpendicular Amp\`ere's laws, which respectively read
\begin{align}   
&\frac{v_A^2}{c^2} \lapp \varphi= \ede N_e + \dd \ds \ede ( \Ppee - N_e) \nno \\
& -(I_0 (2 \dd \ds )\edde -1)  \varphi  \nno \\
&+(I_0 (2 \dd \ds) - I_1 (2 \dd \ds)) \edde \bpar  \nno \\
&- \edi N_i - \tau \ds \edi (\Ppei - N_i)- (I_0 (2 \tau \ds) \eddi -1) \frac{\varphi}{\tau}  \nno \\
&-(I_0 (2 \tau \ds) - I_1 (2 \tau \ds )) \eddi \bpar, \label{poissg}
\end{align}
\beq   \label{amppg}
\begin{split}
\lapp \apar = \frac{\beta_e}{2} ( \ede U_e - \edi U_i ),
\end{split}
\eeq
and
\beq   \label{amppeg}
\begin{split}
& \bpar=- \frac{\beta_e}{2}\left( \ede \Ppee + \dd \ds \ede ( \Ppee -N_e) \right.  \\
& \left. - (I_0 (2 \dd \ds ) - I_1 (2 \dd \ds)) \edde \varphi \right.  \\
& \left. + 2 (I_0 (2 \dd \ds) - I_1 (2 \dd \ds)) \edde \bpar \right.  \\
& \left. + \tau \edi \Ppei + \tau^2 \ds \edi (\Ppei - N_i)\right. \\
&\left.  + (I_0 (2 \tau \ds) - I_1 (2 \tau \ds))\eddi \varphi \right.  \\
& \left. +2 \tau (I_0 (2 \tau \ds) - I_1 (2 \tau \ds)) \eddi \bpar\right).
\end{split}
\eeq
The operators  ${\overline \Gamma}_0$ and ${\overline \Gamma}_1$ and $\Delta_s$ are defined as
%
%
${\overline \Gamma}_0(z,z')=I_0(z z')\exp(z+z')$, 
${\overline \Gamma}_1(z,z')=I_1(zz')\exp(z+z')$ and 
$\Delta_s=\frac{1}{2}\lapp$,
with $I_0$ and $I_1$ indicating the modified Bessel function of the first kind of order zero and one, respectively. 

The set of gyrofluid equations (\ref{neg})-(\ref{amppeg}) was derived in Ref. \cite{Bri92}, although with a different normalization and with the combination $I_0+I_1$ instead of $I_0-I_1$ in Eqs. (\ref{poissg}) and (\ref{amppeg}).
In Eqs. (\ref{neg})-(\ref{amppeg}), we corrected a few typographical errors that were present in the corresponding equations of Ref. \cite{TSP16} (where they had no effect in the considered asymptotics).

\bibliography{biblio}

\end{document}